\documentclass[11pt,a4paper]{article}
\pdfoutput=1

\usepackage{jcappub}

\usepackage{bm}
\usepackage{epsfig}
\usepackage{citesort}
\usepackage{graphicx}
\usepackage{amsmath}
\usepackage{amssymb}
\usepackage{amsbsy}
\usepackage{psfrag}
\renewcommand\[{\left[}

\newcommand{\DS}{ {\scriptscriptstyle \rm  D\hspace{-0.07cm}S} }

\newcommand{\Ref}{Ref.}
\newcommand{\Sec}{Sec.}

\newcommand{\Tabs}{Tabs.}

\newcommand{\fig}{Fig.}

\newcommand{\bea}{\begin{eqnarray}}
\newcommand{\eea}{\end{eqnarray}}

\newcommand{\be}{\begin{equation}}
\newcommand{\ee}{\end{equation}}

\newcommand{\ba}{\begin{array}}
\newcommand{\ea}{\end{array}}

\newcommand{\ie}{\emph{i.e.}}
\newcommand{\eg}{\emph{e.g.}}

\newcommand{\Neff}{N_{\rm eff}}
\newcommand{\exclude}[1]{}

\begin{document}
\subheader{\hfill CERN-PH-TH/2012-070; MPP-2012-56}

\title{Asymmetric Dark Matter and Dark Radiation}

\author[a]{Mattias~Blennow}
\author[b]{Enrique~Fernandez~Martinez}
\author[c]{Olga~Mena}
\author[d]{Javier~Redondo}
\author[e]{Paolo~Serra}

\affiliation[a]{Max-Planck-Institut f\"ur Kernphysik, Saupfercheckweg 1, 69117 Heidelberg,
Germany}
\affiliation[b]{CERN Physics Department, Theory Division, CH-1211 Geneva 23, Switzerland}
\affiliation[c]{Instituto de F\'{\i}sica Corpuscular, CSIC-Universitat de Val\`encia, \\ 22085, E-46071 Valencia, Spain}
\affiliation[d]{Max-Planck-Institut f\"ur Physik %(Werner-Heisenberg-Institut)
F\"ohringer Ring 6, D-80805 M\"unchen, Germany}
\affiliation[d]{Institut d'Astrophysique Spatiale, UMR8617,
Universite Paris-Sud \& CNRS, Bat. 121, Orsay F-91405, France}

\emailAdd{Mattias.Blennow@mpi-hd.mpg.de}
\emailAdd{enfmarti@cern.ch}
\emailAdd{omena@ific.uv.es}
\emailAdd{redondo@mppmu.mpg.de}
\emailAdd{serra@astron.nl}

\abstract{
Asymmetric Dark Matter (ADM) models invoke a particle-antiparticle asymmetry, similar to the one observed in the Baryon sector, to account for the Dark Matter (DM) abundance. Both asymmetries are usually generated by the same mechanism and generally related, thus predicting DM masses around 5~GeV in order to obtain the correct density. The main challenge for successful models is to ensure efficient annihilation of the thermally produced symmetric component of such a light DM candidate without violating constraints from collider or direct searches. A common way to overcome this involves a light mediator, into which DM can efficiently annihilate and which subsequently decays into Standard Model particles. Here we explore the scenario where the light mediator decays instead into lighter degrees of freedom in the dark sector that act as radiation in the early Universe. While this assumption makes indirect DM searches challenging, it leads to signals of extra radiation at BBN and CMB. Under certain conditions, precise measurements of the number of relativistic species, such as those expected from the Planck satellite, can provide information on the structure of the dark sector. We also discuss the constraints of the interactions between DM and Dark Radiation from their imprint in the matter power spectrum.  }

\maketitle

\section{Introduction}
By now we have overwhelming evidence for the presence of an extra non-baryonic Dark Matter (DM) component in the Universe from a variety of different independent sources (see \eg, \Ref~\cite{Bertone:2004pz}). These include rotation curves of galaxies, gravitational lensing, structure formation and global fits of cosmological data such as the cosmic microwave background (CMB) anisotropies measured by the WMAP satellite. However, all this evidence comes exclusively from gravitational effects and we remain sadly ignorant of the properties characterizing the new particle species constituting the DM, such as their masses or the types and strengths of their interactions with the Standard Model (SM) particles. Even in the case where DM is composed by a single stable species, it is a reasonable expectation that this will be only one ingredient of a richer sector with more particles and interactions, given the complexity of the SM sector. The model-building possibilities are therefore limitless, given the drought of information on the dark sector (DS) we currently suffer. 

In this context, most of the interest over the last years has been focused in obtaining a DM candidate within theories that try to address other shortcomings of the SM. Notable examples are the axion~\cite{Peccei:1977hh}, introduced to solve the strong CP problem, and the popular weakly interacting massive particle (WIMP)~\cite{Ellis:1983ew}, with a $\sim 100$~GeV mass, which can be easily accommodated in models addressing the electroweak hierarchy problem. Indeed, the masses of the extra degrees of freedom required to stabilize the Higgs mass under radiative corrections cannot be much further away from the electroweak scale if the fine tuning problem is to be addressed. The fact that the relic abundance of a weakly interacting particle with $\sim 100$~GeV mass gives the observed DM energy density is an additional bonus for these theories and is usually dubbed the ``WIMP miracle''.

Lately, a paradigm shift in the study of DM models is taking place with the goal of exploring the phenomenological consequences of, not only the standard WIMP and axion DM models, but also those of other plausible dark sectors. After all, while the electroweak hierarchy and the strong CP problems are well-motivated theoretical puzzles on their own, the only solid experimental evidence we have so far for physics beyond the SM is DM ---along with neutrino masses and mixings--- and it is fully justified to develop SM extensions with only the aim of addressing the existence and nature of DM.
This is an important avenue to pursue in order to prevent our experimental efforts from becoming too focused on the leading DM paradigms, possibly missing relevant phenomenological signals. 
In this context, the old idea of asymmetric dark matter~(ADM)~\cite{Nussinov:1985xr,Barr:1990ca,Barr:1991qn,Kaplan:1991ah,Kuzmin:1996he,Kusenko:1998yi} is becoming increasingly popular~\cite{Farrar:2004qy,Hooper:2004dc,Kitano:2004sv,Agashe:2004bm,Kitano:2005ge,Cosme:2005sb,Farrar:2005zd,Suematsu:2005kp,Tytgat:2006wy,Banks:2006xr,Page:2007sh,Kitano:2008tk,Nardi:2008ix,Kaplan:2009ag,Kribs:2009fy,Cohen:2009fz,Cai:2009ia,An:2009vq,Frandsen:2010yj,An:2010kc,Cohen:2010kn,Taoso:2010tg,Shelton:2010ta,Davoudiasl:2010am,Haba:2010bm,Belyaev:2010kp,Chun:2010hz,Buckley:2010ui,Gu:2010ft,Blennow:2010qp,Hall:2010jx,Dutta:2010va,Falkowski:2011xh,Haba:2011uz,Chun:2011cc,Heckman:2011sw,Graesser:2011wi,Frandsen:2011kt,McDermott:2011jp,Buckley:2011kk,Iminniyaz:2011yp,Bell:2011tn,Cheung:2011if,Davoudiasl:2011fj,MarchRussell:2011fi,Cui:2011qe,Arina:2011cu,Buckley:2011ye,Barr:2011cz,Cirelli:2011ac,Lin:2011gj,Petraki:2011mv,vonHarling:2012yn,Kamada:2012ht,Iocco:2012wk}. ADM offers a DM paradigm in which the origin and properties of DM are much more closely related to those of baryonic matter. This seems appealing, since both abundances are observed to be close to each other $\Omega_{DM} \approx 5 \Omega_{b}$. 
Indeed, if the origin and mass of the DM candidate are similar to the baryons, this coincidence is less striking than within the ``WIMP miracle'', where the production mechanism and masses in the dark and visible sectors are completely different. ADM models postulate that the stability of the DM population stems from a new conserved quantum number, $X$. The relic density is then associated to a particle-antiparticle asymmetry, in complete analogy to the baryonic sector and baryon number. 
A common origin for both the baryon and DM asymmetries is usually assumed, which typically implies similar abundances and, therefore, a constraint on the DM mass close to 5~GeV so as to reproduce the correct observed energy density. This precise prediction turns out to be quite general if the mechanism linking the DM and baryon asymmetries conserves a combination of $B-L$ and $X$, say $Q$. In this case, the DM mass turns out to be $5-7/Q_{\rm DM}$, with $Q_{\rm DM}$ the $Q$-charge of the dark matter particle~\cite{Ibe:2011hq}. The main challenge of successful ADM models is to provide sufficient annihilation of the thermally produced symmetric component with such a light DM candidate without violating collider or direct search constraints~\cite{Buckley:2011kk,MarchRussell:2012hi}. The most common solution involves a lighter mediator with $\sim 10-100$~MeV mass~\cite{Lin:2011gj} in the DS, into which DM can efficiently annihilate and which subsequently decays into SM particles. This lighter mediator would be the DS analogue of the pion, which leads to efficient annihilation of the symmetric component in the baryon sector.
  
In this work we will take the analogy between the dark and visible sectors one step further and assume that the DS, in addition to the DM and the mediator, contains very light degrees of freedom which, like photons and neutrinos in the SM sector, contribute to the radiation content of the Universe. We will dub this content ``dark radiation'' (DR). Since the symmetric component of DM and any mediators can now annihilate or decay into DR and not into SM particles, the connection between the DS and the SM weakens, alleviating the constraints stemming from 
collider, DM direct and indirect detection experiments\footnote{However, if the mediator itself constitutes the DR, the long range DM-DM interactions implied are strongly constrained through structure formation and can be ruled out in many scenarios~\cite{Lin:2011gj}.  
Similar constraints have been studied in the context of non-ADM models~\cite{Ackerman:2008gi,Feng:2008mu,Feng:2009mn}, such as scenarios with a hidden sector photon mediating DM-DM interactions~\cite{Ackerman:2008gi}.}.  
Also, in models where the symmetry related to the DM number $X$ is explicitly violated by a small Majorana mass term, the DM interactions with the DR bath could play an important role in DM$\leftrightarrow$anti-DM oscillations, which can normally challenge the survival of the asymmetry~\cite{Buckley:2011ye,Cirelli:2011ac,Tulin:2012re}.

This scenario offers two new probes into the structure of the DS, which are complementary to direct, indirect and collider DM searches. 
On one hand, after the SM and DS decouple, the thermal symmetric populations of DS particles end up annihilating via the light mediator into DR, which becomes heated with respect to the photon thermal bath. 
Thus, constraints on the allowed amount of extra radiation can lead to constraints on the DS degrees of freedom, providing valuable information for model building. The amount of energy in dark radiation is traditionally expressed in terms of the extra effective number of neutrinos $\Delta \Neff $ that can be probed by its effect on the cosmic microwave background (CMB) and the outcome of big bang nucleosynthesis (BBN). 
Interestingly, these estimates of $\Delta \Neff $ presently show a trend towards a non-zero DR component~\cite{Mangano:2006ur,Hamann:2007pi,Reid:2009nq,Dunkley:2010ge,Komatsu:2010fb,Hamann:2010pw,Keisler:2011aw,Archidiacono:2011gq,Smith:2011ab,Hamann:2011hu,Nollett:2011aa,Izotov:2010ca,Aver:2011bw,Aver:2010wq} which shall be either confirmed or excluded by the great improvement in sensitivity expected in the near future from the CMB data of the Planck satellite.  
On the other hand, since some interaction between the ADM population and the DR must exist via the lighter mediator, this interaction can be bounded through constraints on the matter power spectrum. Indeed, for significant DM-DR interactions, DM and DR can form a coupled fluid which, in a way analogous to the photon-baryon plasma, is not pressureless and can propagate acoustic oscillations. 
We will discuss the constraints that galaxy surveys can set on this DM-DR interactions.

The present study on the presence of extra relativistic components in the DS that interact at some level with the DM population has been inspired by the asymmetric DM paradigm, in which complex dark sectors with extra light degrees of freedom are generally required. However, we will try to keep our study as model independent as possible and hence our results can also apply to different DM models, not necessarily based in a particle-antiparticle asymmetry, in which DM interacts with a DR component. Indeed, the constraints on $\Delta \Neff $ stemming from BBN and the CMB analysis that we will discuss can also apply to other models of DM. In principle this is also true for the bounds on DM-DR interactions that we will derive from the matter power spectrum. However, as we will see, the size of the interactions required to lead to any observable effect rules out that DM is a thermal relic. Indeed, the annihilation of DM to DR would be too large to reproduce the observed DM abundance. The ADM paradigm, on the other hand, decouples the DM abundance, determined by a particle-antiparticle asymmetry, from its annihilation cross section and could thus lead to the signals we will constrain. Indeed, large annihilation cross sections are particularly desirable in ADM models so as to efficiently remove the thermal symmetric DM component that can otherwise dominate over the asymmetry and spoil its relation to the baryon abundance.

This paper is organized as follows: In \Sec~\ref{sec:DR}, we derive the constraints that present and near future measurements of $\Delta \Neff $ imply on the degrees of freedom present in the dark sector as a function of the decoupling temperature. Section~\ref{sec:DR-DM} is dedicated to reviewing the bounds that can be derived on the interactions between the DM and DR populations from the galaxy power spectrum. Finally, in \Sec~\ref{sec:summary}, we summarize our results and give our conclusions.

\section{Dark Radiation}
\label{sec:DR}

If the thermal component of DM ends up annihilating into DR, an extra contribution to the energy density in  relativistic degrees of freedom will be present, $\rho_{\rm DR}$. This contribution is generally parametrized through $\Delta \Neff $, the number of extra effective neutrino species by normalizing the contribution of the extra radiation to that of a neutrino field 
\begin{equation}
\Delta \Neff  = \frac{\rho_{\rm DR}}{2\frac{7}{8}\frac{\pi^2}{30}T_\nu^4} , 
\end{equation}
where $T_\nu$ is the temperature of neutrinos at the specific moment of interest.

Many different possibilities can be envisioned for the thermal histories of the SM and the DS depending on the details of particular DM realizations, the structure of the DS, and its interactions with the SM. 
A common feature of ADM models is that they usually contain significantly more structure than the light stable field constituting the DM. Furthermore, given the flavour structure observed in the SM, it seems naive to assume that the DS, which amounts to a five times larger fraction of the energy content of the Universe, would actually be only composed by a single field. In this context, models with flavoured DM are becoming increasingly popular~\cite{Ibanez:1983kw,Hagelin:1984wv,Freese:1985qw,Falk:1994es,Servant:2002aq,Blennow:2010qp,Kile:2011mn,Batell:2011tc,Cui:2011qe,Kamenik:2011nb,Agrawal:2011ze}.
In this work, we will adopt an approach as model independent as possible. We parametrize the potential complexity of the DS through the number of relativistic degrees of freedom $g_h+g_\ell$ present in the DS at the time of decoupling from the SM, characterized by a temperature $T_d$. While $g_\ell$ corresponds to the number of degrees of freedom present in the field(s) that ultimately constitute the DR, the $g_h$ degrees of freedom correspond to relatively heavy degrees of freedom that are going to turn non-relativistic and heat the DR with respect to the SM. This general approach implies that the results of this section can apply, not only to ADM models by which we were inspired, but also to any DS that contains light degrees of freedom which constitute DR apart from the DM candidate.

Two extreme examples of our parametrization are the following: If only the heaviest fields in the DS interact with the SM, $T_d$ will be very high and most of the DS degrees of freedom will be relativistic at decoupling, allowing arbitrarily high values of $g_h$ depending on the complexity of the model. The opposite scenario would be realized when the DR fields have interactions with the SM which keep them in thermal equilibrium until very late times. In this case, $g_h = 0$ and the DR will not be heated again with respect to the photon bath and will have a final lower temperature depending on the moment of decoupling. This last example can be realized if DR is made up of sterile neutrinos that are mixed with the SM ones and decouple almost at the same time.

The SM and DS share a common temperature until $T_d$ and from that point onwards they evolve independently.  In this case, the comoving entropies of the two sectors are conserved separately. We can use this fact to track the temperature changes in each sector, which we need to evaluate $\Delta \Neff $.
The comoving entropy in a thermalized sector is defined as
\begin{equation}
S= \frac{2 \pi^2}{45}g^s_{*} T^3 a^3,
\end{equation}  
where $T$ is the common temperature of the sector, $a$ is the scale factor, and $g^s_{*}$ is the \emph{effective number of entropy degrees of freedom}. The latter can be conveniently expressed as a sum over species 
\begin{equation}
g^s_{*}(T)= \sum_{i=\rm bosons} g_i f^-_i + \frac{7}{8}\sum_{j=\rm fermions} g_i f^+_j 
\end{equation} 
where $g_i$ is the number of internal degrees of freedom of species $i$ and 
\begin{equation}
f_i^\pm = \frac{45}{4 \pi^4}\left(\frac{8}{7}\right)^{\frac{1\pm 1}{2}} z_i^4\int_1^\infty \frac{y\sqrt{y^2-1}}{\exp (y z_i)\pm 1}\frac{4 y^2-1}{3 y}dy
\end{equation} 
are functions of the particle mass $m_i$ through the ratio $z_i=m_i/T_x$ that changes smoothly from $f^\pm_i=1$ when the particle is relativistic ($z_i\ll 1$) to $f^\pm_i=0$ when it becomes non-relativistic ($z_i\gg1$). In practice we can consider them as a filter that allows only relativistic species to contribute to $g^s_*$. For the SM $g^s_*(T)$ we have used the approximate fitted expression of Appendix A of Ref.~\cite{Wantz:2009it}, which takes into account its non-trivial behavior around the QCD phase transition (which cannot be reproduced with the simple formulas above). 
From now on, we will not use any subindex for SM quantities ($g^s_*,T$) and subindex D\hspace{-0.1cm}S for DS quantities ($g^s_{*,\DS},T_{\DS}$).

At this stage, it is convenient to develop a general formula for the temperature difference of two sectors.  After decoupling the comoving entropies are constant during the expansion, and so it is their ratio, 
$S_1/S_2=g^s_{*,1}T_1^3/g_{2,*} T_2^3= {\rm constant}$, where all these four quantities can depend on time through the temperatures. The constant can be evaluated at the time of decoupling, where $T_1=T_2=T_d$, to be $S_1/S_2=g^s_{*,1}(T_d)/g^s_{*,2}(T_d)$. 
The ratio of the two temperatures at any later time is therefore
\begin{equation}
\label{T1T2}
\frac{T_1}{T_2} = \left(\frac{g^s_{*,1}(T_d)}{g^s_{*,1}} \frac{g^s_{*,2}}{g^s_{*,2}(T_d)}\right)^{1/3} ,  
\end{equation}
where the $g^s_{*i}$ and $T_i$ are understood to be evaluated at the same time.

We will now apply this formula to compute $\Delta \Neff $ as a function of the DS degrees of freedom. 
It is convenient to distinguish two different cases.

%%%%%%% %%%%%%% %%%%%%% %%%%%%% %%%%%%% %%%%%%%
\subsection*{High DS decoupling temperature, $T_d\gtrsim$ MeV} 
%%%%%%% %%%%%%% %%%%%%% %%%%%%% %%%%%%% %%%%%%%

The simplest and probably more realistic case occurs when the decoupling temperature of the hidden sector is higher than the neutrino-electron decoupling, i.e. $T_d \gtrsim$ MeV. 
In terms of the light and heavy DS degrees of freedom introduced before,  
the relative temperature between DR and photons at the CMB epoch is then given by:\\
\begin{equation}
\left.\frac{T_\DS}{T_\gamma}\right|_{\rm CMB} = \left( \frac{(g_h+g_l)}{g_l}\frac{g^s_{*\rm CMB}}{g^s_*(T_d)} \right)^{1/3},
\end{equation}
with $g^s_{*\rm CMB}$ being the SM effective number of entropy degrees of freedom at the CMB epoch. 
This number receives contributions from photons and neutrinos, but since the latter ones decouple before 
the electron/positron annihilation they do not share the corresponding entropy injection and have a smaller temperature. The neutrino/photon temperature ratio follows from Eq.~\ref{T1T2} by encompassing photons and electrons/positrons in sector 2, $T_\nu/T_\gamma=(4/11)^{1/3}$. It follows that $g^s_{*\rm CMB}=2 \left( 1 + \frac{7}{8}\frac{4}{11}3 \right) \simeq 3.909$. 
{It is worth noting that when electrons and positrons annihilate, neutrinos are not fully decoupled and electroweak reactions are able to pump a bit of energy into the neutrino sector. 
This makes the standard value of $N_{\rm eff}\simeq 3.046$~\cite{Mangano:2005cc}. 
This $\sim 2\%$ correction it is too small to be significantly detected even by the best prospects of Planck, it teaches us however that 
incomplete thermalization or decoupling of species in the DS can lead to non-integer values of $g_h,g_l$.}

If we assume that all the DS light degrees of freedom have the same final temperature, the energy density is $\rho_{\rm DR}=\pi^2 g_l T_\DS^4/30$, and we find 
\begin{equation}
\left.\Delta \Neff \right|_{\rm CMB} = \frac{g_l }{2\frac{7}{8} \left(\frac{4}{11} \right)^{4/3}} 
\left(\left.\frac{T_\DS}{T_\gamma}\right)^4\right|_{\rm CMB}=  \frac{13.56}{ g^s_*(T_d)^{4/3}} \frac{(g_l+g_h)^{4/3}}{g_l^{1/3}}.
\label{eq:dof}
\end{equation}
It is easy to prove that in this scenario the value of $\Delta \Neff $ measured through BBN, $\left.\Delta \Neff \right|_{\rm BBN}$, is equal or smaller than $\left.\Delta \Neff \right|_{\rm CMB}$. 
Before getting into it, we recall that the definition of $\left.\Delta \Neff \right|_{\rm BBN}$ differs from that of $\left.\Delta \Neff \right|_{\rm CMB}$. 
The time interval where $\left.\Delta \Neff \right|_{\rm BBN}$ affects BBN spans from the decoupling of the beta reactions that keep protons and neutrons in chemical equilibrium at high temperatures ($T\sim$ MeV) to proper BBN times (end of the deuterium bottleneck, $T\sim 70$ keV). 
As we already mentioned, between these two boundaries, the electron/positron annihilation heats photons but not neutrinos, changing the neutrino to photon temperature ratio and therefore the definition of $\Delta \Neff$. 
It is customary to define $\left.\Delta \Neff \right|_{\rm BBN}$ evaluated at the highest temperature $T\sim$ MeV and this is the definition we use in this work.   
      
At this epoch  photons, electrons and neutrinos have the same temperature and $g^s_{*\rm BBN}=10.75$ and following the same steps as before we obtain
\begin{equation}
\left.\Delta \Neff \right|_{\rm BBN} = \frac{g_l^{\rm BBN}}{2\frac{7}{8}} 
\left(\left.\frac{T_\DS}{T_\gamma}\right)^4\right|_{\rm BBN}=  \frac{13.56}{ g^s_*(T_d)^{4/3}} \frac{(g_l+g_h)^{4/3}}{(g_l^{\rm BBN})^{1/3}}.
\end{equation}
where $g_l^{\rm BBN}$ is the number of relativistic degrees of freedom in the DS at the time of BBN, which may take on values between $g_l$ and $g_l + g_h$.
%we have explicitly extracted from $g_h$ the DS degrees of freedom that are still relativistic at BBN times but will end up decaying or annihilating into light degrees of freedom before the CMB epoch, $g_m$. 
If $g_l^{\rm BBN}=g_l$, then $\Delta \Neff $ has the same value for CMB and BBN physics, but if larger, the BBN value is smaller. 
It is interesting to note that by fixing $g_l+g_h$ and $T_d$ we are \emph{fixing} the comoving entropy of the DS, but the final number of degrees of freedom $g_l$ (or $g_l^{\rm BBN}$ for BBN) still has an impact on $\Delta \Neff $. 
Actually, the dependence $\Delta \Neff  \propto (g_l)^{-1/3}$ is quite easy to understand. The reason is that, for a fixed entropy density $s\propto g T^3$, the energy density $\rho \propto g T^4\propto s^{4/3}/g^{1/3}$ is larger for larger temperatures and small number of degrees of freedom. 
This implies that, for two sectors with the same number of degrees of freedom at decoupling, that one with the smallest (non-zero) number of light species will be more noticeable to $\Delta \Neff $ probes. 

Finally, note that in this scenario we can be sensitive to $g_l^{\rm BBN}-g_l$ by measuring and comparing $\left.\Delta \Neff \right|_{\rm BBN}$ and $\left.\Delta \Neff \right|_{\rm CMB}$. Unfortunately, the foreseeable accuracy of  $\left.\Delta \Neff \right|_{\rm BBN}$ makes this comparison challenging. 

%%%%%%% %%%%%%% %%%%%%% %%%%%%% %%%%%%% %%%%%%%
\subsection*{Low DS decoupling temperature, $T_d\lesssim$ MeV} 
%%%%%%% %%%%%%% %%%%%%% %%%%%%% %%%%%%% %%%%%%%

In this case, the DS has to couple either to the electromagnetic plasma (electrons, baryons and photons) or to neutrinos, which at these low temperatures are decoupled from each other. Here we do not consider the case that the DS can mediate interactions between neutrinos and the electromagnetic sector. 
Having a DS in thermal contact with the SM below the MeV requires very strong interactions between the two, so the reader should keep in mind that many of the models that we can constrain in this section can actually be excluded by laboratory experiments or astrophysical arguments. 
In this case, the DS is still coupled during BBN so computing $\left.\Delta \Neff \right|_{\rm BBN}$ amounts to just counting the DS degrees of freedom present at $T\sim$ MeV and normalize them to the neutrino energy density 
\begin{equation}
\left.\Delta \Neff \right|_{\rm BBN}=\frac{4}{7}(g_H+g_h+g_l)  ,  
\end{equation}
where $g_H$ is the number of degrees of freedom that become non-relativistic between the BBN and the DS decoupling. We  are forced to introduce this new parameter in order to maintain the meaning of $g_h$ and $g_l$ as in the case discussed above. 

Let us now compute $\left.\Delta \Neff \right|_{\rm CMB}$. In the case in which the DS remains coupled to neutrinos until $T_d$ we have
\begin{equation}
\left. \frac{T_\nu}{T_\gamma}\right|_{T_d}=
\left. \frac{T_\DS}{T_\gamma}\right|_{T_d}=
\left(\frac{3\times 7/4+g_H+g_h+g_l}{3\times 7/4+g_h+g_l}\frac{2}{2+7/2}\right)^{1/3} , 
\end{equation}
where $21/4$ are the neutrino degrees of freedom, and we have again assumed $T_d<m_e$. When the DS decouples from neutrinos, it can still get heated with respect to them by the usual factor $T_\DS/T_\nu=((g_h+g_l)/g_l)^{1/3}$. In this case we find 
\begin{equation}
\left.\Neff \right|_{\rm CMB} = 
\left(3+\frac{4}{7}\frac{(g_h+g_l)^{4/3}}{(g_l)^{1/3}}\right)
\left(\frac{3\times 7/4+g_H+g_h+g_l}{3\times 7/4+g_h+g_l}\right)^{4/3} . 
%\label{eq:dof}
\end{equation}

If, on the other hand, the DS couples preferentially to the electromagnetic sector (ES) we have $T_\DS=T_\gamma=T_d$ at decoupling. 
Neutrinos are decoupled from $T\sim$ MeV, where the ES degrees of freedom are $2+7/2$ for photons and $e^\pm$. Thus at $T_d$ they have a temperature
\begin{equation}
\left. \frac{T_\nu}{T_\gamma}\right|_{T_d}=
\left. \frac{T_\nu}{T_\DS}\right|_{T_d}=\left(\frac{2+g_h+g_l}{2+7/2+g_H+g_h+g_l}\right)^{1/3},
\end{equation}
where we have assumed $T_d\ll m_e\sim$ MeV so that only photons are present at the DS decoupling. 
Also we have used that $g^s_{*,\nu}($MeV)$=g^s_{*,\nu}(T_d)$. 
Below $T_d$, also the DS decouples and later reduces its degrees of freedom from $g_h+g_l$ to $g_l$ before the CMB epoch.  In this period the DS temperature increases with respect to the photon one by a factor $T_\DS/T_\gamma=((g_h+g_l)/g_l)^{1/3}$. Since, in this case, the neutrinos do not have their standard temperature ratio to photons, it is more convenient to quote the number of effective neutrino species, and not only the \emph{extra} component ($\Delta \Neff =\Neff -3$)
\begin{equation}
\left.\Neff \right|_{\rm CMB} = 
3\left(\frac{11}{4}\frac{2+g_h+g_l}{2+7/2+g_H+g_h+g_l}\right)^{4/3}+g_l\frac{4}{7}\left(\frac{11}{4}\frac{g_h+g_l}{g_l}\right)^{4/3} . 
%\label{eq:dof}
\end{equation}
Let us note that this scenario suffers from an additional constraint. The entropy contained in the $H$ species is dumped into the ES, between the BBN and CMB epochs. This entropy ends up in photons at the CMB epoch, but is absent during BBN and therefore the baryon-to-photon ratio $\eta$, the parameter that governs the output of BBN will be different between these two epochs,  
\begin{equation}
\frac{\eta_{\rm BBN}}{\eta_{\rm CMB}} = \frac{2+g_h+g_l}{2+g_H+g_h+g_l} . 
\end{equation}
The relic abundance of deuterium is quite sensitive to the value of $\eta_{\rm BBN}$, and can be therefore used to constraint any mismatch between $\eta_{\rm CMB}$ and $\eta_{\rm BBN}$, see for instance~\cite{Jaeckel:2008fi,Cadamuro:2011fd,Cadamuro:2010cz}. 
This is because the current BBN prediction using $\eta_{\rm BBN}=\eta_{\rm CMB}=(6.19\pm 0.15)\times 10^{-10}$~\cite{Nakamura:2010zzi} already agrees with the observations within the uncertainties~\cite{Nakamura:2010zzi}. 
A nonzero $g_H$ will make $\eta_{\rm BBN}$ smaller than $\eta_{\rm CMB}$, increasing the Deuterium yield for a fixed  $\eta_{\rm CMB}$. In our scenarios, we also have a nonzero $\Delta \Neff $, but actually this also predicts an increase in D/H so both effects go in the same direction.  
We have checked that indeed these arguments exclude values other than $g_H=0$. 

%%%%%%% %%%%%%% %%%%%%% %%%%%%% %%%%%%% %%%%%%%
\subsection*{Bounds, hints and forecasts}
%%%%%%% %%%%%%% %%%%%%% %%%%%%% %%%%%%% %%%%%%%

We can now use present data on $\Delta \Neff $ to constrain the structure of the hidden sector, \ie, $g_H, g_h$ and $g_l$. At present, constraints from CMB and BBN physics are similar but not too restrictive. 
There seems to exist a trend in favor of a non-zero value of $\Delta \Neff >0$, which could be hinting at the kind of more complex dark sectors which we study in this work. 

Let us first consider BBN, which has so far provided the strongest constraints on $\Delta\Neff$~\cite{Iocco:2008va,Pospelov:2010hj}.  
An increased value of $\Neff$ leads to a faster expansion of the Universe and, as consequences of this, 
larger Helium mass fraction $Y_p$ (since neutrons have higher freeze-out abundance and have less time to decay before BBN), larger Deuterium to proton ratio D/H (because D burning reactions are less effective) and smaller yields of more massive nuclei like Lithium (because they are produced from D and its products, whose reactions are slower). 
At present, comparisons of the observed primordial abundances with the theoretical predictions show a trend towards $\Delta \Neff>0$, with best fit values $\Delta \Neff\sim 0.5-0.8$~\cite{Izotov:2010ca,Mangano:2011ar,Hamann:2010bk,Hamann:2011ge}. 
Interestingly, both the Deuterium and the Helium measurements show this preference.  
The observed primordial abundance of Lithium is much smaller than the predictions from standard cosmology. This is the so-called \emph{Lithium problem}, which can be somewhat alleviated by the presence of extra radiation.  
However, the cited works conclude that these preferences are not significant given the errors, and in particular the systematic uncertainties involved in the estimation of the primordial abundances. 
Notwithstanding the above, one can obtain robust upper bounds on $\Delta \Neff$ like $\Delta \Neff\leq 1$~\cite{Mangano:2011ar} and $\Delta \Neff\leq 1.26$~\cite{Hamann:2011ge} (both at 95\% C.L.). 
These bounds can be relaxed in non-minimal scenarios, as for instance if neutrinos have a nonzero chemical potential (up to $\Delta \Neff\leq 2.56$ at 95~\%~C.L.\ according to \Ref~\cite{Hamann:2011ge}). 
Thus, it seems that BBN does not contradict the presence of extra radiation, but cannot be used to assess it quantitatively nor to exclude it model independently beyond the $\Delta \Neff\leq 2.56$ level. 
As we explain in the following, the situation can be much different for the CMB. 

Many claims for an excess $\Delta \Neff  \sim 1$ during the CMB release are present in the literature~\cite{Mangano:2006ur,Hamann:2007pi,Reid:2009nq,Dunkley:2010ge,Komatsu:2010fb,Hamann:2010pw,Keisler:2011aw,Archidiacono:2011gq,Smith:2011ab} with different levels of significance depending on the dataset and analysis performed in the study. While it has been shown that these effects can be amplified by volume effects when analyzing data with Bayesian statistics and the effect seems to be significantly prior dependent for some datasets~\cite{GonzalezMorales:2011ty}, the preference for non-zero $\Delta \Neff $ does persist for prior-independent frequentist analyses~\cite{Hamann:2011hu}. Here we will take the latest results from the South Pole Telescope collaboration in combination with WMAP data~\cite{Keisler:2011aw} as a reference $\Delta \Neff  = 0.85 \pm 0.62$.

It should be noted that the presence of extra radiation during matter-radiation equality to which the CMB is sensitive does not necessarily imply its presence during BBN. For this reason, we will regard them as independent constraints into two different epochs of the early Universe and not combine them. 
In this work we will mainly focus on the constraints from CMB probes. Indeed, the CMB not only offers a window to a lower temperature to which DR must contribute even if it was not present at BBN, but also the forthcoming results from the Planck mission will soon provide much more stringent constraints superseding present BBN sensitivity. Note that, in order to reach its full potential, Planck data should in any case be combined with BBN results on the primordial He abundance $Y_p$~\cite{Hamann:2007sb}.

\begin{figure}
\begin{center}
\includegraphics[width=.60\textwidth]{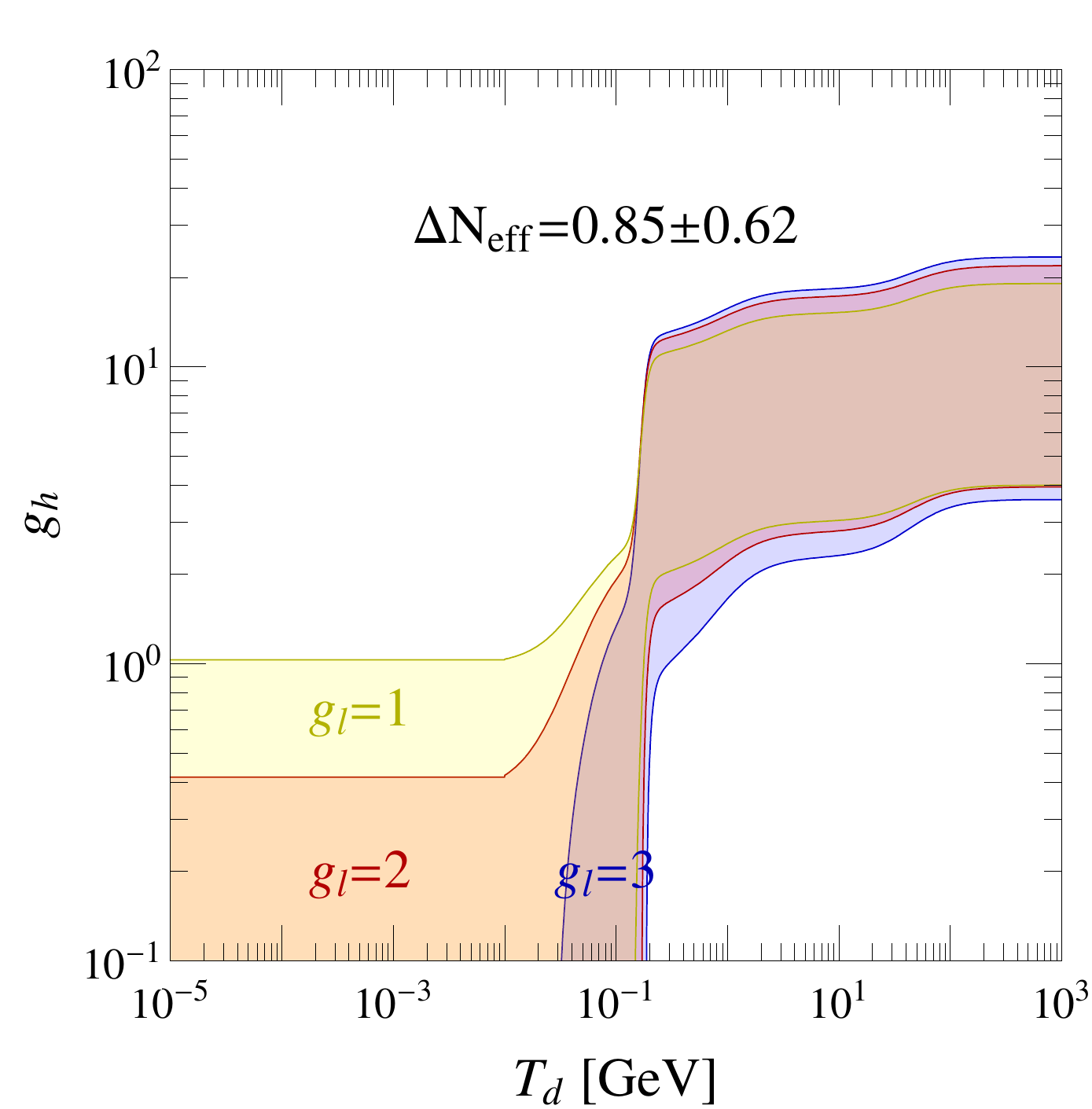}
\caption{The $1 \sigma$ range for the number of heavy degrees of freedom $g_h$ required to heat the light sector in order to account for the presently preferred number of extra effective neutrino species during CMB $\Delta \Neff = 0.85 \pm 0.62$ and as a function of the decoupling temperature $T_d$.}
\label{fig:dNeffpresent}
\end{center}
\end{figure}
In \fig~\ref{fig:dNeffpresent} we show the constraints on $g_h$ as a function of $T_d$ for $\Delta \Neff  = 0.85 \pm 0.62$ at CMB~\cite{Keisler:2011aw} and a given $g_\ell$. Below 1~MeV the results correspond to the (less constrained) scenario where the DS remains coupled to neutrinos after BBN and $g_H = 0$. For $g_H \neq 0$, similar results are obtained, but interpreting $g_h$ in the vertical axis as $g_h + g_H$. As can be seen from this figure, for very late decoupling, having extra heavy degrees of freedom is increasingly disfavored. This is due to the fact that the photon bath will not receive significant heating from the SM sector at such low temperatures and thus, any extra heating in the DS would lead to a too large contribution to $\Delta \Neff$. On the other hand, if the SM decouples from the DS at a higher temperature, then the relativistic degrees of freedom in the SM will be heated, requiring heating also in the DS in order for it to contribute significantly to $\Delta \Neff $. 

The Planck satellite mission is expected to measure the effective number of neutrino species with an excellent accuracy. We compute as a first step the CMB Fisher matrix to obtain forecasts for the Planck satellite~\cite{:2006uk}. Our fiducial model is a $\Lambda$CDM  cosmology with five parameters: the physical baryon and CDM densities, $\omega_b$ and $\omega_{\rm DM}$, the scalar spectral index, $n_{s}$, $h$ (being the Hubble constant $H_0=100\ h$~km Mpc$^{-1}$s$^{-1}$) and the dimensionless amplitude of the primordial curvature perturbations, $A_{s}$ (see Tab.~\ref{tab:fiducial_standard_model} for their values). Furthermore, we add to the $\Lambda$CDM fiducial cosmology a number of DR degrees of freedom parametrized as extra sterile neutrino species $\Delta \Neff =1,2,3$. 
We assume that the sterile species have thermal spectra and are not coupled among themselves. 
For these fiducial cosmologies, our Fisher forecast analysis provides the following errors: $\Delta \Neff =1 \pm 0.08$,  $\Delta \Neff =2 \pm 0.08$ and $\Delta \Neff =3 \pm 0.1$ at $1\sigma$. We then refine this analysis and perform a Markov Chain Monte Carlo simulation of the expected Planck results when the total number of neutrinos is 3, 4, 5 or 6, which correspond to the cases $\Delta \Neff =0,1,2$ and $3$ respectively. For the Monte Carlo scan we obtained good agreement with the Fisher matrix results: $\Delta \Neff  < 0.08$, $\Delta \Neff =1 \pm 0.10$,  $\Delta \Neff =2 \pm 0.11$ and $\Delta \Neff =3 \pm 0.14$ at $1\sigma$. Therefore, near future data from Planck will definitely be able to settle the issue of DR. If evidence for $N_\nu>3$ still persists after these new accurate CMB measurements, it will be extremely interesting to further study interacting scenarios in the DR and DM sectors.

%%%%%%%%%%%%%%%%%%%%%%%%%%%%%%%%%%%%%%%%%%%%%%%%%%%

\begin{table}[htbp]
\centering
\begin{tabular}{c|c|c|c|c|c}
\hline\hline
$\Omega_bh^2$ & $\Omega_{\rm DM}h^2$ & $n_{s}$ & $h$ & $A_{s}$&$\Delta \Neff $\\
0.02267&0.1131&0.96&0.705&$2.64\cdot 10^{-9}$&1-3\\
\hline\hline
\end{tabular}
\caption{Values of the parameters in the fiducial models explored in this study.}
\label{tab:fiducial_standard_model}
\end{table}
%%%%%%%%%%%%%%%%%%%%%%%%%%%%%%%%%%%%%%%%%%%%%%%%%%%

%
\begin{figure}
\begin{center}
\includegraphics[width=.48\textwidth]{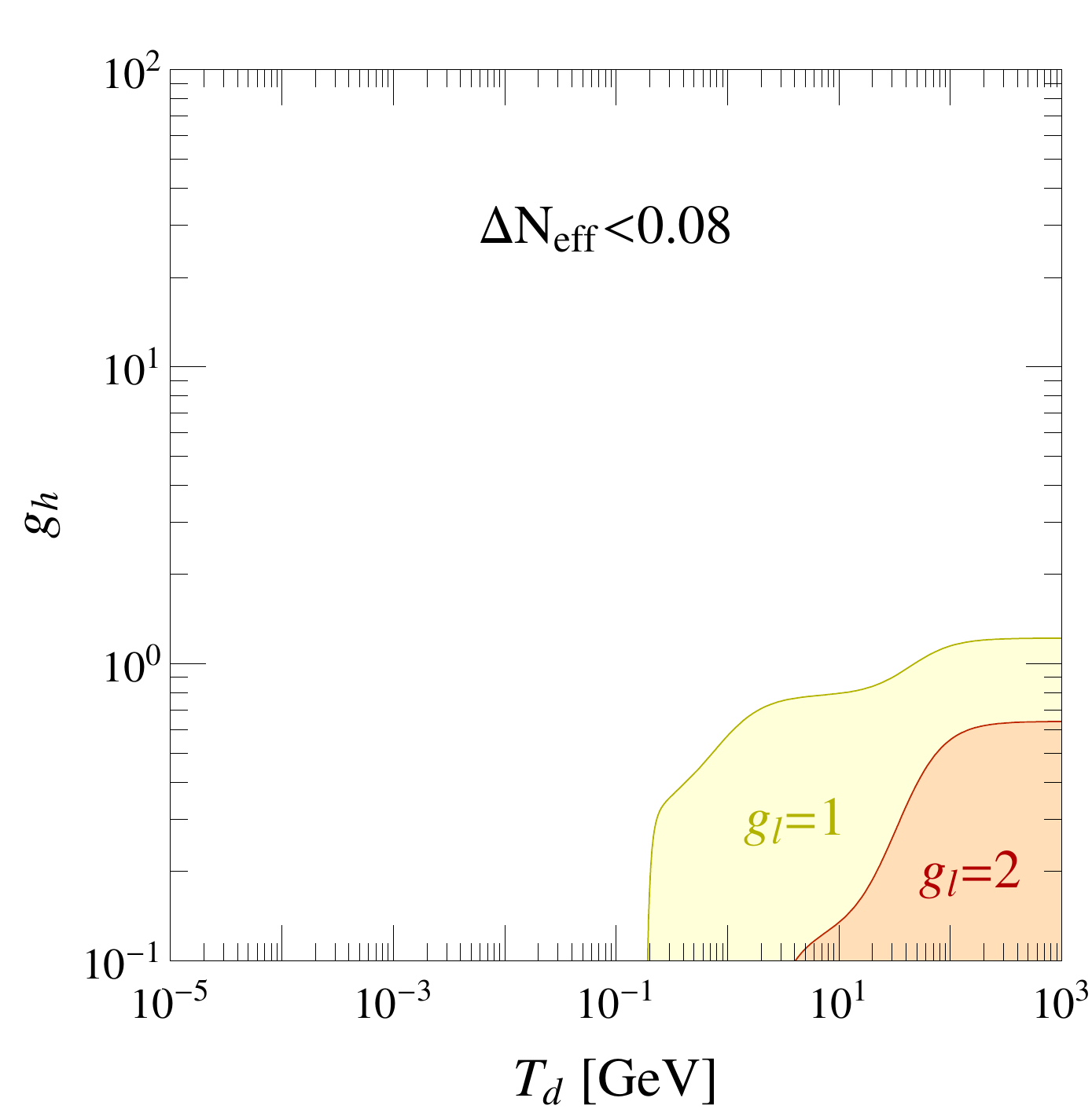}
\includegraphics[width=.48\textwidth]{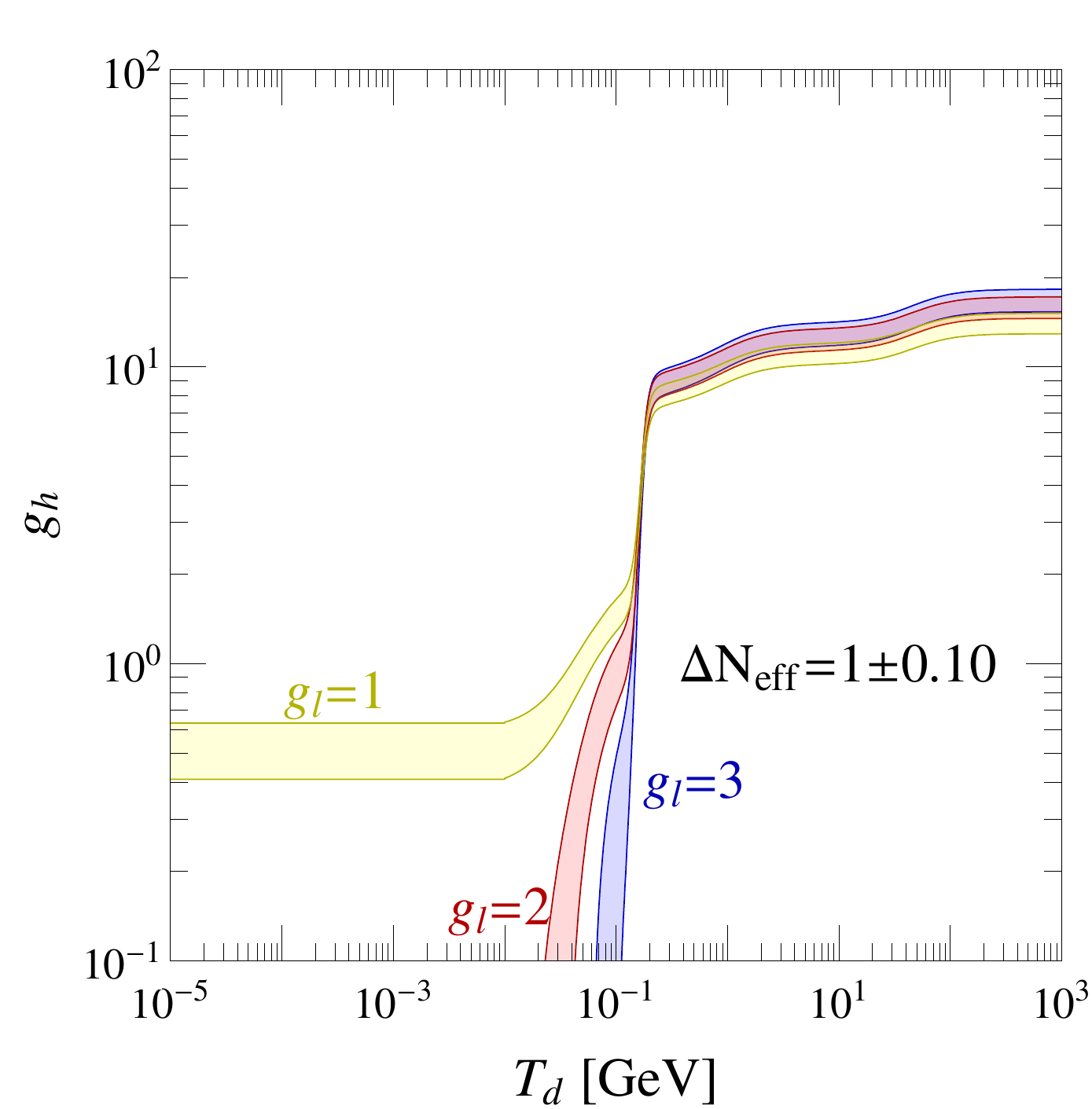} \\
\includegraphics[width=.48\textwidth]{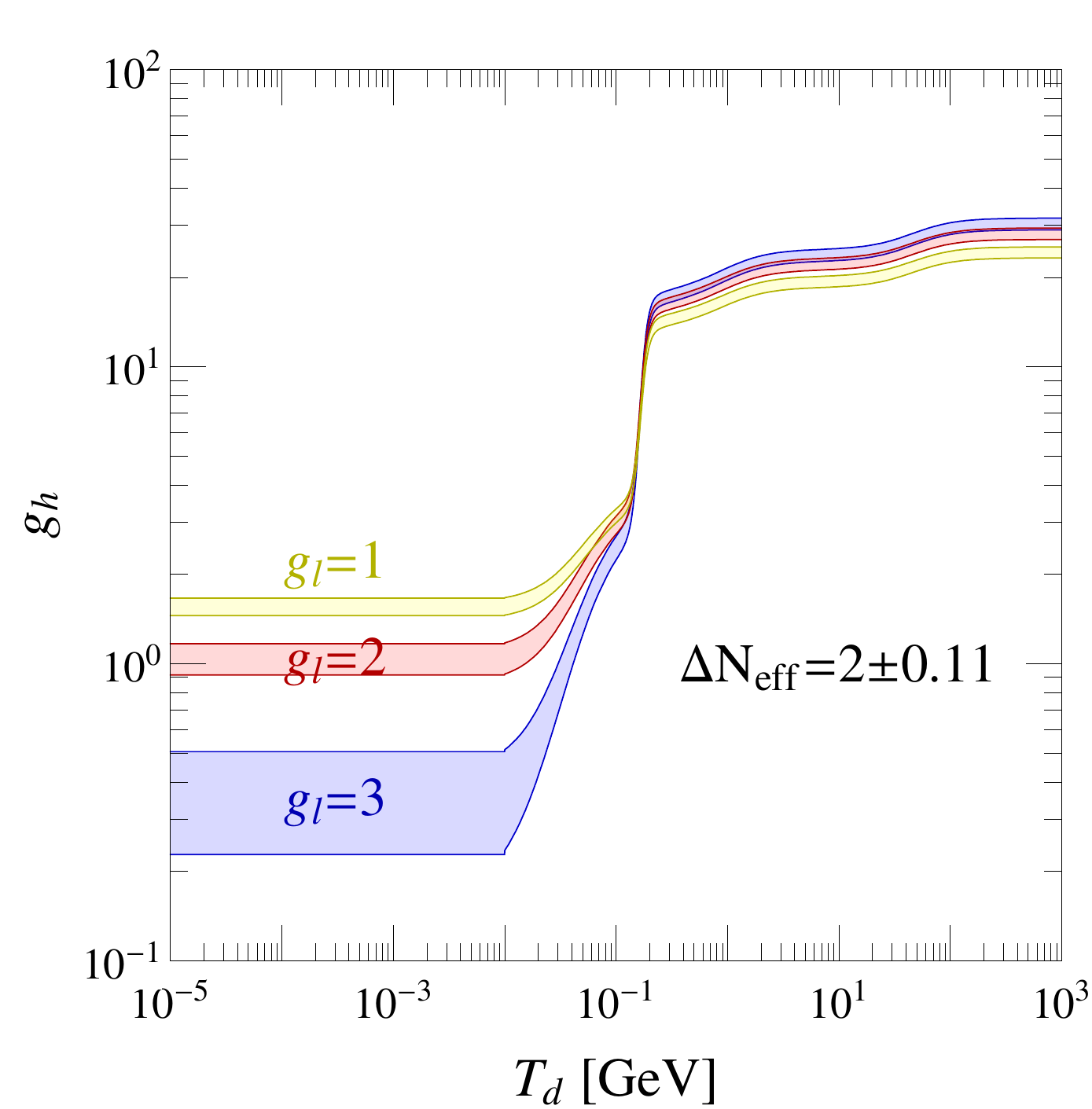}
\includegraphics[width=.48\textwidth]{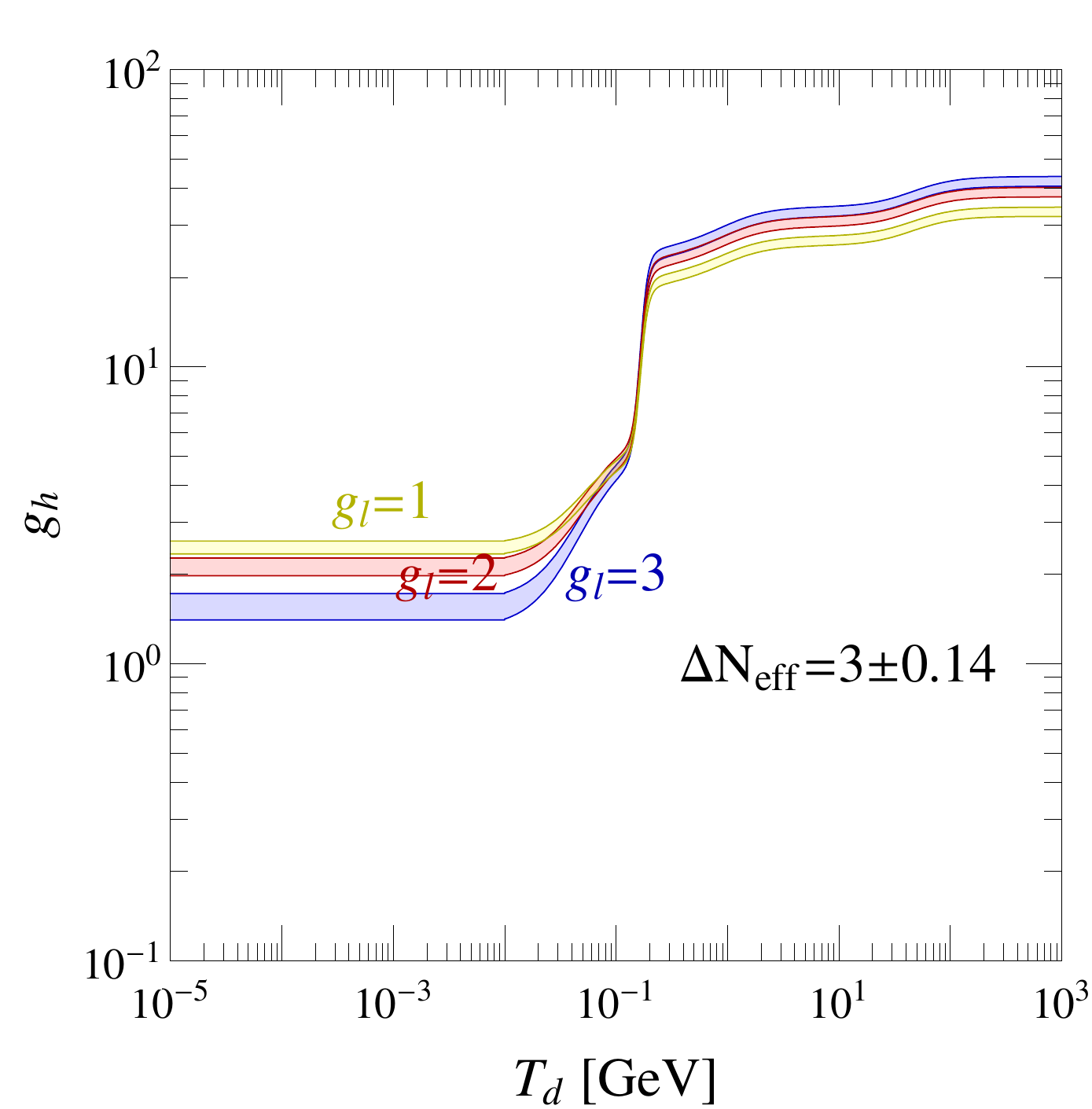} \\
\caption{The $1 \sigma$ range for the number of heavy degrees of freedom $g_h$ required to heat the light sector in order to account for a given number of extra effective neutrino species during CMB and as a function of the decoupling temperature $T_d$. The bands correspond to different Planck constraint forecasts as described in the text.}
\label{fig:dNeff}
\end{center}
\end{figure}
For the different scenarios that Planck could find, we show in \fig~\ref{fig:dNeff} the vast improvement that the smaller errors would imply in the constraints on $g_h$ and, thus, on the complexity of the DS. Notice that, in the extreme scenario in which Planck data would prefer 3 extra effective neutrino species, some heating from $g_h$ is required even at decoupling temperatures as low as 1~MeV and for $g_l=3$. This extreme scenario is significantly disfavoured by present BBN and CMB results, though. In the opposite limit in which Planck finds no evidence for extra radiation and only sets an upper limit on $\Delta \Neff $, no contours at the $1 \sigma$ level are allowed for $g_l=3$ for any decoupling temperature, as long as the SM is only heated by the SM particle content with respect to the dark sector after decoupling. 

%%%%%%%%%%%%%%%%%%%%%%%%%%%%%%%%%%%%%%%%%%%%%%%%%%%
\section{Dark Matter-Dark Radiation interactions}
%%%%%%%%%%%%%%%%%%%%%%%%%%%%%%%%%%%%%%%%%%%%%%%%%%%
\label{sec:DR-DM}

The interactions between the DR and DM components will leave an imprint on the galaxy power spectrum, see Refs.~\cite{Mangano:2006mp,Serra:2009uu}. In the presence of these interactions, the dark matter fluid is no longer pressureless and therefore the situation will be analogue to that of baryons and photons before the recombination era,  with a series of damped oscillations similar to the baryon acoustic oscillations. 
In analogy to the baryon case, DM-DR interactions will modify the matter power spectrum at scales similar to the size of the Universe when they become ineffective in changing the velocities of DM particles, after which DM particles begin to fall into potential wells.  
We denote this typical length scale by $1/k_f \sim 1/a_{f}H (a_f)$ where $H=d\log a/dt$ is the expansion rate of the universe and the scale factor at freeze-out $a_f$ is approximately given by solving
\begin{equation}
H(a_f)= \Gamma (a_f) = \frac{\rho_{\rm DR}}{\rho_{\rm DM}}n_{\rm DM} \left< \sigma_{\rm DM-DR} v \right>
= \frac{\left<  E_{\rm DR} \right>}{m_{\rm DM}} n_{\rm DR} \left< \sigma_{\rm DM-DR} v \right>
\end{equation}
where $\Gamma$ is the rate at which DM velocities are changed by O(1) amounts, $\rho$ and $n$ denote the energy and number density of dark matter or dark radiation, $m_{\rm DM}$ is the DM mass and $\left<  E_{\rm DR} \right>$ the average energy of DR particles.  

\begin{figure}[h]
\begin{center}
\begin{tabular}{c} 
\includegraphics[width=12cm]{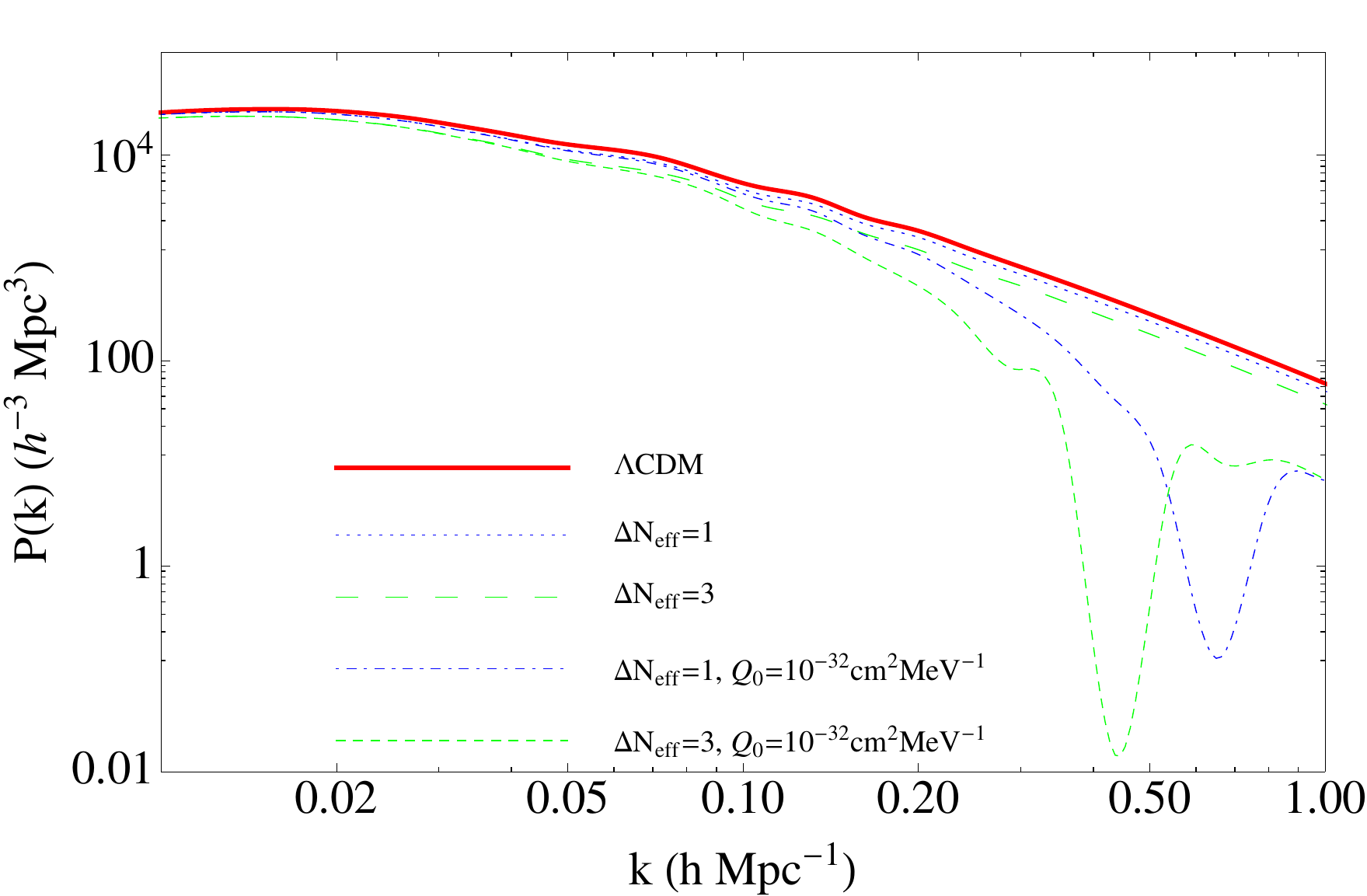}\\
\includegraphics[width=12cm]{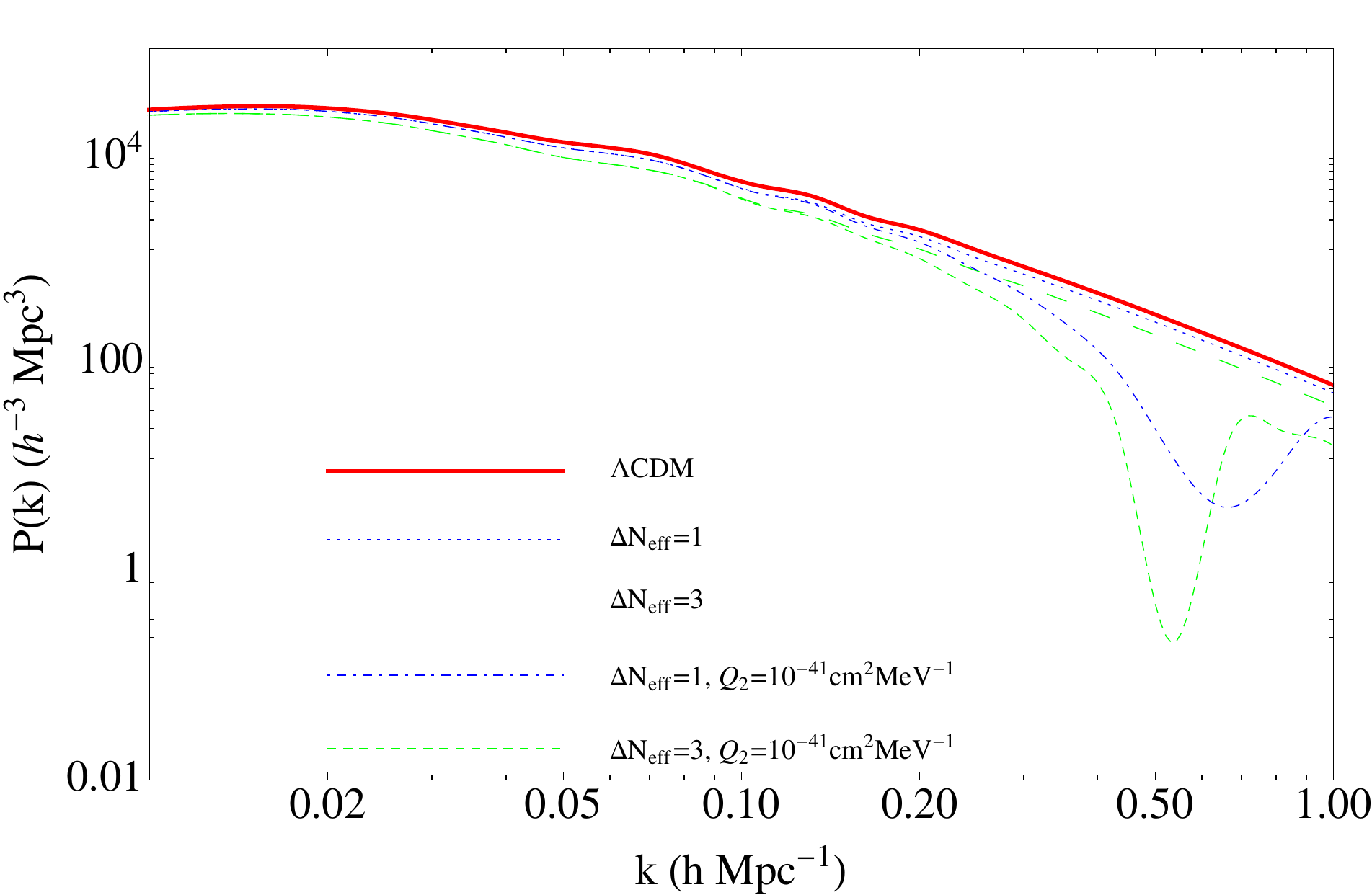}
\end{tabular}
\end{center}
\caption{Upper panel: Matter power spectrum for a $\Lambda$CDM model (thick red curve). The blue (dotted) dotted-dashed lines depict the matter power spectrum for $\Delta \Neff =1$ within a (non) interacting scenario with constant cross section and $Q_0=10^{-32}$~cm$^2$~MeV$^{-1}$. The green (long) short dashed lines depict the matter power spectrum for  $\Delta \Neff =3$ within a (non) interaction scenario. The lower panel shows the analogous but for an interaction cross section $\propto 1/a^2$ and  $Q_2=10^{-41}$~cm$^2$~MeV$^{-1}$.}

\label{fig:matterpower}

\end{figure}

It is convenient to parameterize the cross section as
\begin{equation}
\left< \sigma_{\rm DM-DR} v \right> = Q_0 \ m_{\rm DM} ~,  
\end{equation}
if it is constant or
\begin{equation}
\left< \sigma_{\rm DM-DR} v \right> = \frac{Q_2}{a^2} \ m_{\rm DM}~, 
\end{equation}
if proportional to $T^2$~\cite{Mangano:2006mp}. 
Here $Q_0$ and $Q_2$ are constants with units cm$^2$~MeV$^{-1}$. 
These two cases are representative of possible DR-DM interactions and help to study how the possible temperature dependence of the cross section can affect the constraints.  
The typical scale of the DM-DR oscillations in these cases are  
\begin{equation}
\label{eq:scale1}
k\sim 0.5 \left(\frac{10^{-32}~{\rm cm}^2~{\rm MeV}^{-1}}{Q_0}\right)^{1/2}~h\textrm{Mpc}^{-1}~, 
\end{equation}
\begin{equation}
\label{eq:scale2}
k\sim 0.6\left(\frac{10^{-41}~{\rm cm}^2~{\rm MeV}^{-1}}{Q_2}\right)^{1/2}~h\textrm{Mpc}^{-1}~, 
\end{equation}
where we have assumed $\Delta \Neff =3$. The dependence of these scales on $\Delta \Neff$ is however quite mild, increasing $k$ as the number of effective neutrino species decreases. 
Figure~\ref{fig:matterpower} (upper panel) illustrates this effect for a constant interaction cross section, $Q_0=10^{-32}$~cm$^2$~MeV$^{-1}$, for $\Delta \Neff=1$ and $\Delta \Neff=3$ interacting with the DM sector. 
As a comparison, we show as well the shape of the matter power spectrum if these species were non interacting. 
Note that the scale at which the damped oscillations appear is well predicted by the approximated expression given by Eq.~(\ref{eq:scale1}).  The right panel of Fig.~\ref{fig:matterpower} illustrates the analogous but for a cross section $\propto T^2$ and $Q_2=10^{-41}$~cm$^2$~MeV$^{-1}$. Again, the suppression scale is very well approximated by Eq.~(\ref{eq:scale2}).  In the following, we shall exploit the dark matter suppression effects to set bounds on the interacting rates $Q_0$ and $Q_2$ using galaxy clustering data combined with other cosmological datasets.

%%%%%%%%%%%%%%%%%%%%%%%%%%%%%%%%%%%%%%%%%%%%%%%%%%%%%%%%%%%

%%%%%%%%%%%%%%%%%%%%%%%%%%%%%%%%%%%%%%%%%%%%%%%%%%%
%%%%%%%%%%%%%%%%%%%%%%%%%%%%%%%%%%%%%%%%
We have modified the Boltzmann CAMB code \cite{Lewis:1999bs} incorporating the interacting scenarios and extracted cosmological parameters from current data using a Monte Carlo Markov Chain (MCMC) analysis based on the publicly available MCMC package \texttt{cosmomc} \cite{Lewis:2002ah}. We consider here a flat $\Lambda$CDM universe with $\Delta\Neff$ DR species interacting with the dark matter. The scenario is described by a set of cosmological parameters
\begin{equation}
 \label{parameter}
      \{\omega_b,\omega_{\rm DM}, H_0,
       n_s, A_s, Q_0 (Q_2) \}~,
\end{equation}
where $\omega_b\equiv\Omega_bh^{2}$ and $\omega_{\rm DM}\equiv\Omega_{\rm{DM}}h^{2}$ are today's ratios of the physical baryon and cold dark matter densities to the critical density, $H_0$ is the current value of the Hubble parameter, $n_s$ is the scalar spectral index, $A_{s}$ is the amplitude of the primordial spectrum, and $Q_0$, $Q_2$  encode the DM-DR interactions. Our basic data set is the seven--year WMAP CMB data \cite{Larson:2010gs}  (temperature and polarization) with the routine for computing the likelihood supplied by the WMAP team. We analyze the WMAP data together with the luminous red galaxy clustering results from SDSS II (Sloan Digital Sky Survey)~\cite{Reid:2009xm}, with a prior on the Hubble constant from HST (Hubble Space Telescope)~\cite{Riess:2009pu}, including  to these data sets Supernova Ia Union Compilation 2 data~\cite{Amanullah:2010vv}.

Our main results are reported in Tabs.~\ref{tab:Q0bounds} and \ref{tab:Q2bounds}, where we list the constraints on the interaction cross section parameters $Q_0$ and $Q_2$ in three possible interaction scenarios, $\Delta \Neff=1,2,3$. Notice that, particularly in the more common case of the cross section scaling with $T^2$, these constraints are too mild to be significant for the thermal abundance of DM. Thus, while the procedure followed is general, the effects studied are mainly relevant for ADM scenarios, in which larger than thermal cross sections are required to efficiently annihilate the symmetric component. The bounds are stronger as the number of effective neutrino increases, since the typical scale $k$ at which damped oscillations appear in the matter power spectrum decreases with $\Neff$. The maximum $k$ used in the analysis of the matter power spectrum is  $\sim 0.15~h$Mpc$^{-1}$, since effects that appear at larger scales cannot be easily constrained. Also, note that the constraints found in this study are milder than those found by the authors of \cite{Mangano:2006mp,Serra:2009uu} since in their case the species interacting with the dark matter fluid were the active neutrinos and not extra radiation species. 
There exists a large degeneracy between the number of extra radiation species and the dark matter energy density. 
One of the main effects of $\Delta \Neff $ comes from the change of the epoch of the radiation matter equality, and consequently, from the shift of the CMB acoustic peaks. The position of acoustic peaks is given by the so-called acoustic scale $\theta_A$, which reads
\begin{equation}
\theta_A=\frac{r_s(z_{rec})}{r_\theta(z_{rec})}~,
\end{equation}
where $r_\theta (z_{rec})$ and $r_s(z_{rec})$ are the comoving angular diameter distance to the last scattering surface and the sound horizon at the recombination epoch $z_{rec}$, respectively. 
Although $r_\theta (z_{rec})$ almost remains the same for different values of $\Delta \Neff$, $r_s(z_{rec})$ becomes smaller when $\Delta\Neff$ is increased. 
Thus the positions of acoustic peaks are shifted to higher multipoles $\ell$ (smaller angular scales) by increasing the value of $\Delta \Neff$. 
The height of the first acoustic CMB peak will also increase as $\Delta \Neff$ does. Both effects can be compensated by a larger cold dark matter component. Therefore, in our analysis, the cold dark matter component is larger than in the absence of dark radiation species, and the effect in the suppression of the matter power spectrum shown in Fig.~\ref{fig:matterpower} will be less noticeable than in the case in which the interacting species are active neutrinos and extra DR species are absent.

\begin{figure}[!htb]
\centering
\begin{minipage}[t]{0.49\textwidth}
\centering
\includegraphics*[width=\linewidth]{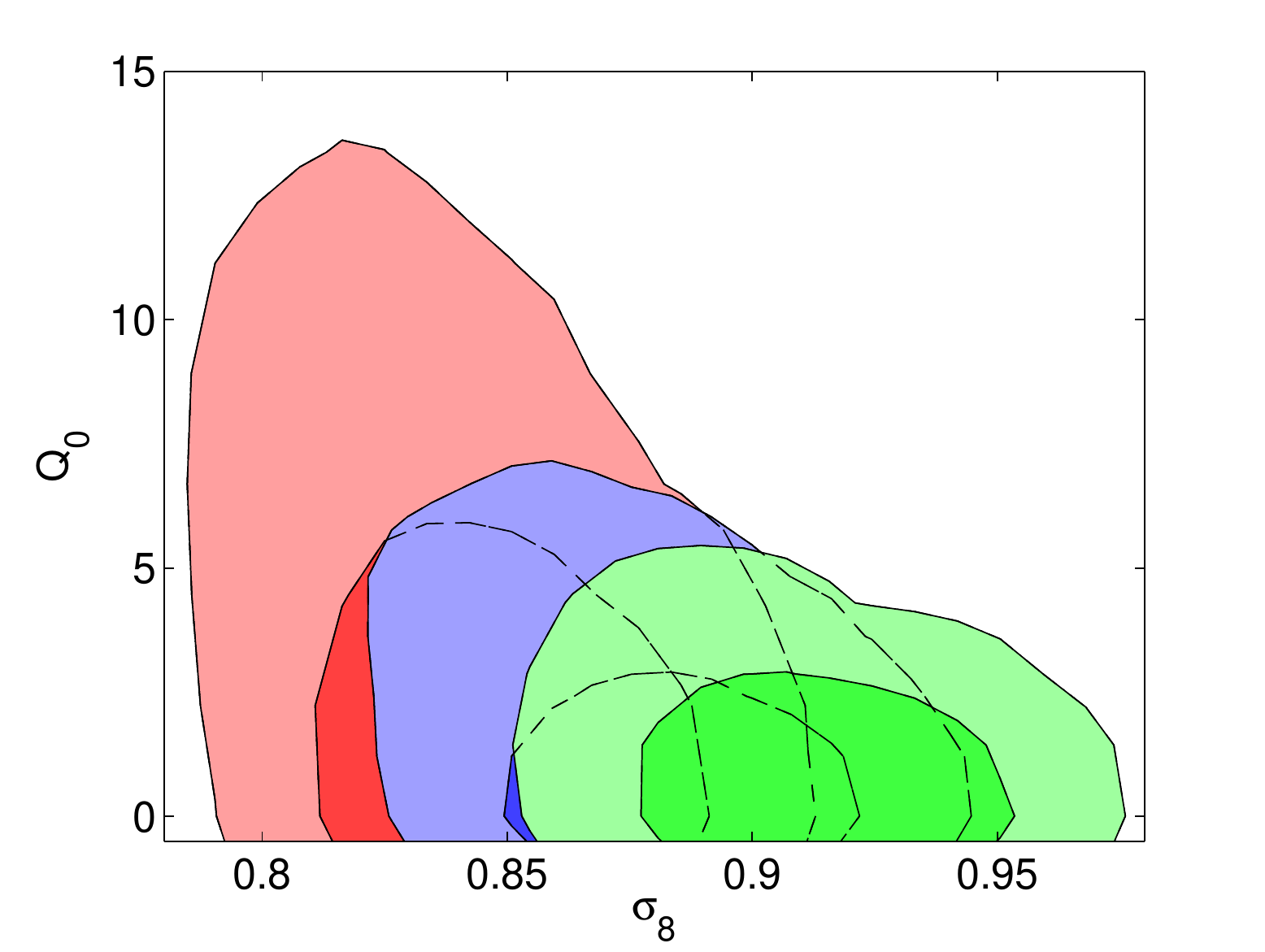}
\end{minipage} \hfill
\begin{minipage}[t]{0.49\textwidth}
\centering
\includegraphics*[width=\linewidth]{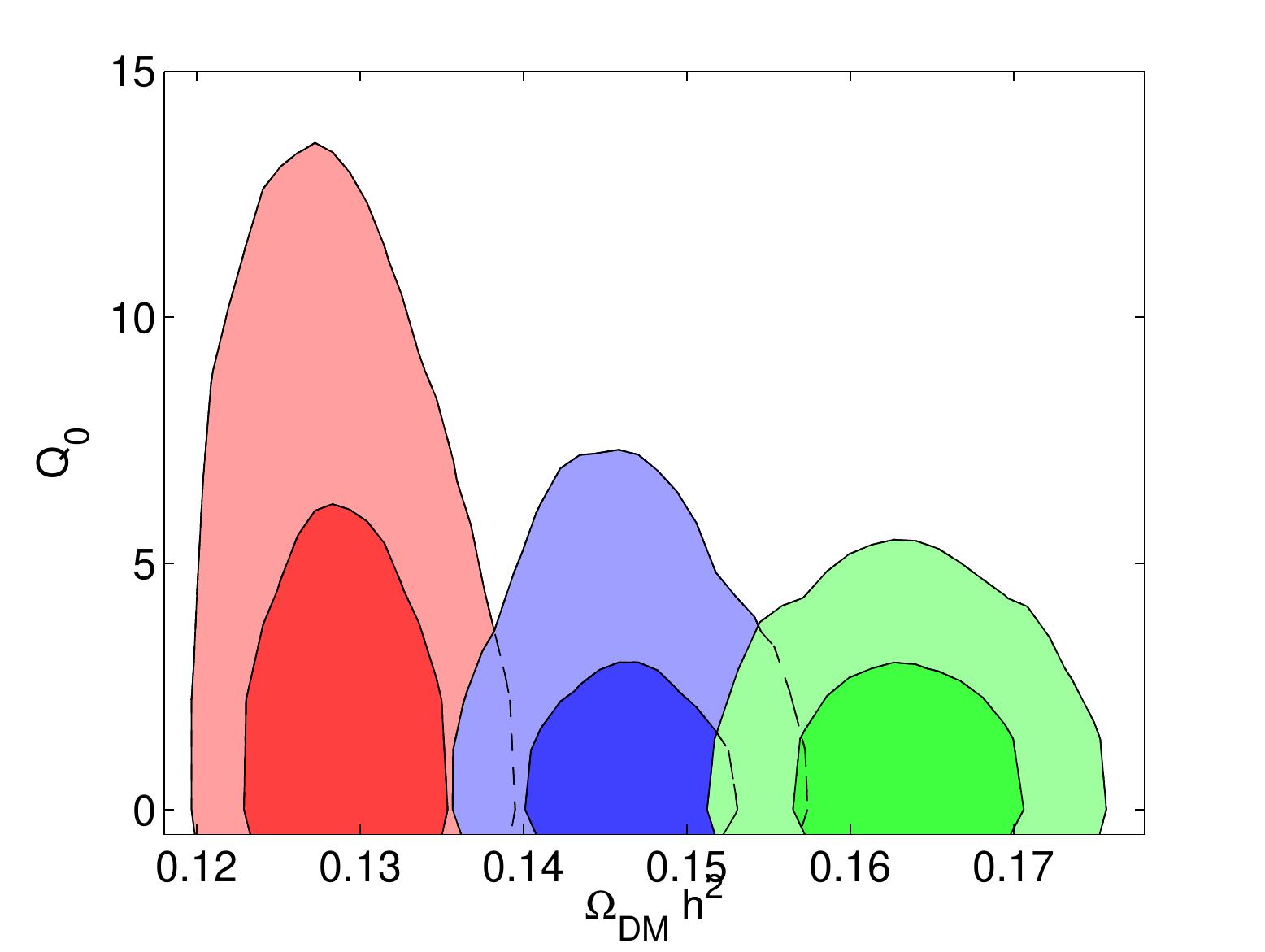}
\end{minipage} \hfill
\begin{minipage}[t]{0.49\textwidth}
\centering
\includegraphics*[width=\linewidth]{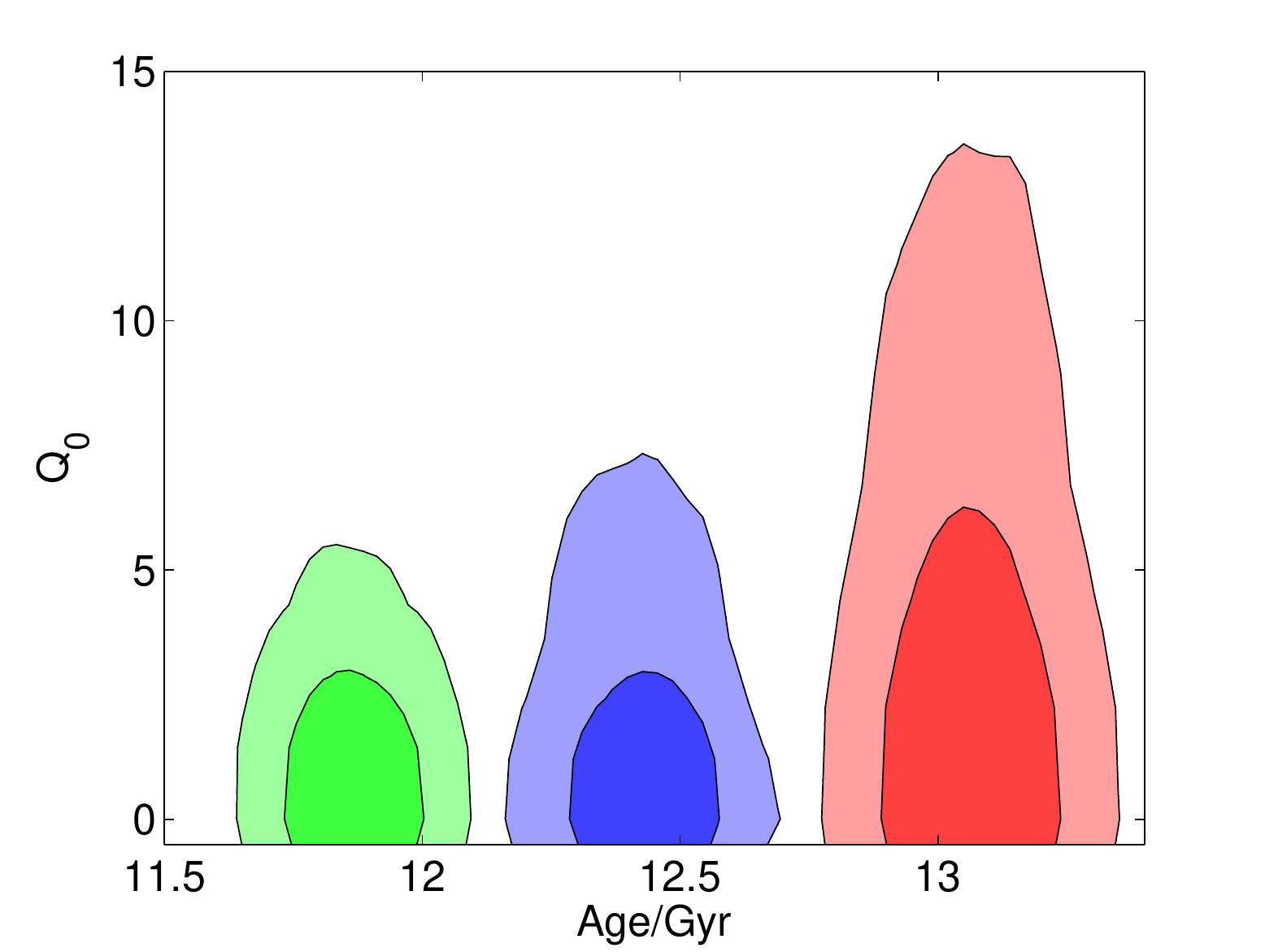}
\end{minipage} \hfill
\caption{The left, right upper panels and the lower panel show the 1$\sigma$ and 2$\sigma$ contours in the ($\sigma_8$, $Q_0$), ($\Omega_{\rm {DM}}h^2$, $Q_0$) and (Age, $Q_0$) planes, respectively. The interacting parameter $Q_0$ is in units of $10^{-34}$~cm$^2$~MeV$^{-1}$ and the age of the universe is in Gyrs. The red, blue and green contours denote the three possible interacting scenarios explored here with one, two and three sterile neutrino species in the DR sector.}
\label{fig:sigma8}
\end{figure}

\begin{figure}[!htb]
\centering
\begin{minipage}[t]{0.49\textwidth}
\centering
\includegraphics*[width=\linewidth]{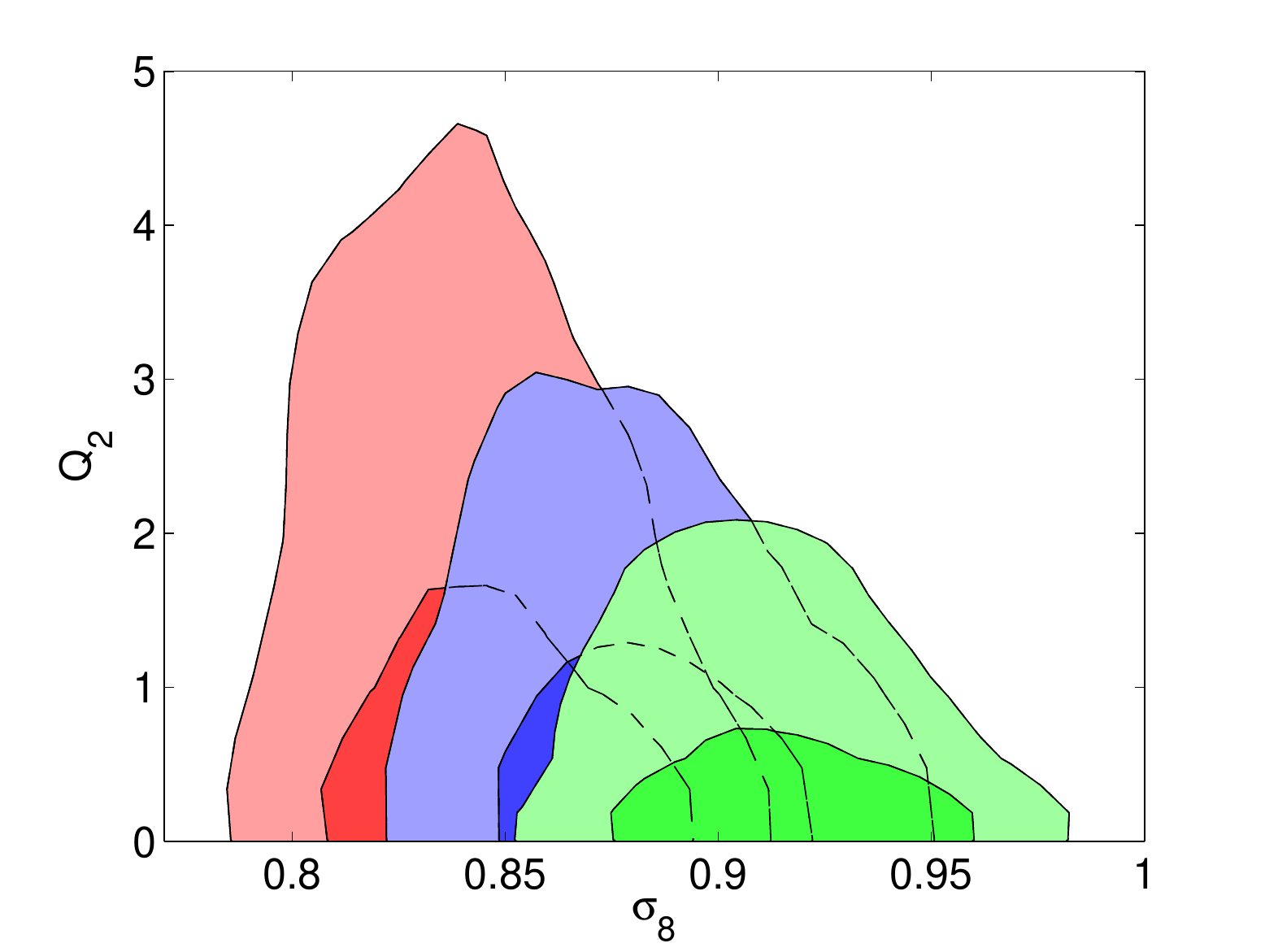}
\end{minipage} \hfill
\begin{minipage}[t]{0.49\textwidth}
\centering
\includegraphics*[width=\linewidth]{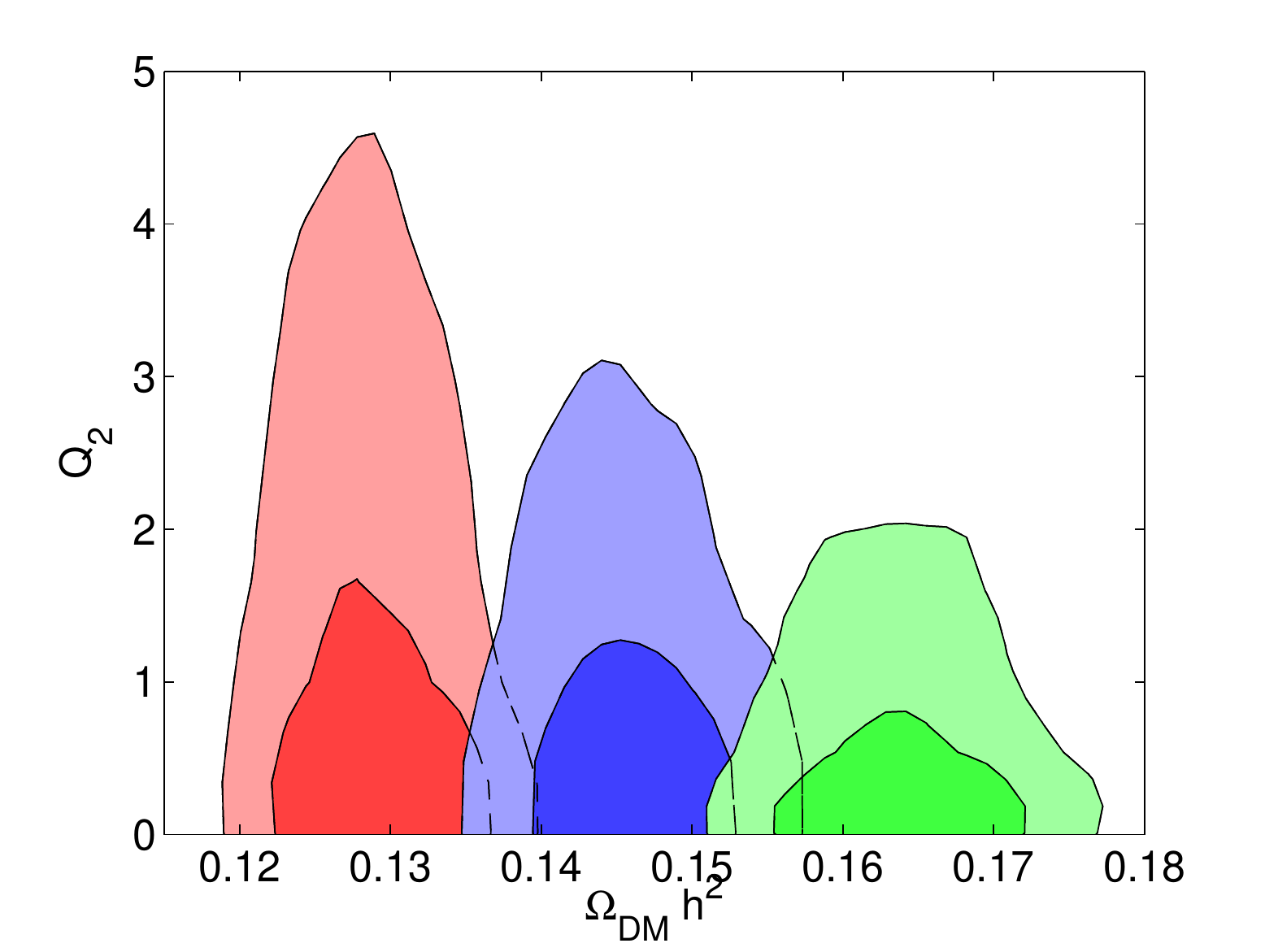}
\end{minipage} \hfill
\begin{minipage}[t]{0.49\textwidth}
\centering
\includegraphics*[width=\linewidth]{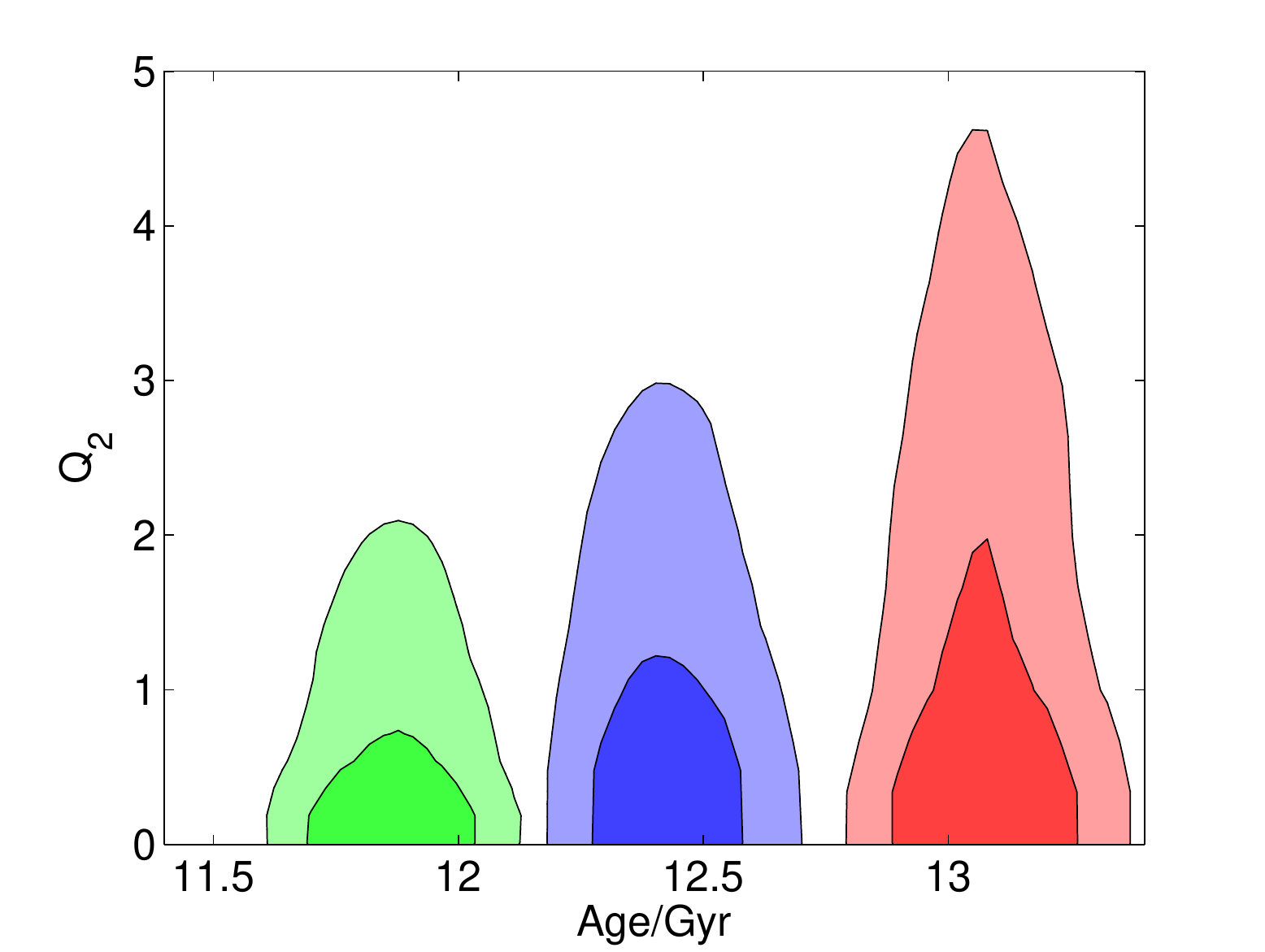}
\end{minipage} \hfill
\caption{The left, right upper panels and the lower panel show the 1$\sigma$ and 2$\sigma$ contours in the ($\sigma_8$, $Q_2$), ($\Omega_{\rm {DM}}h^2$, $Q_2$) and (Age, $Q_2$) planes, respectively. The interacting parameter $Q_2$ is in units of $10^{-43}$~cm$^2$~MeV$^{-1}$ and the age of the universe is in Gyrs. The red, blue and green contours denote the three possible interacting scenarios explored here with one, two and three sterile neutrino species in the DR sector.}
\label{fig:sigma8b}
\end{figure}

Figures \ref{fig:sigma8} and \ref{fig:sigma8b} show the 1$\sigma$ and 2$\sigma$ contours in the ($\sigma_8$, $Q_0$)~\footnote{$\sigma_8$ is defined as the rms matter density fluctuations in spheres of $8$~Mpc.}, ($\Omega_{\rm {DM}}h^2$, $Q_0$) and (Age, $Q_0$) planes, with $Q_0$ and $Q_2$ in units of $10^{-34}$~cm$^2$~MeV$^{-1}$ and $10^{-43}$~cm$^2$~MeV$^{-1}$, respectively, and the age of the Universe in Gyrs. 
The contours are shown for the three possible interaction scenarios considered here, with one, two and three DR species. 
Notice that the scenarios with $\Delta\Neff>1$ are increasingly disfavoured. Indeed, taking the minimum of the log-likelihood of the different scenarios, we observe a difference of $\log(L_{max}(\Delta\Neff=3)/L_{max}(\Delta\Neff=1)) \sim 2$. Although this suggests that the scenario with $\Delta\Neff=1$ is more favored by data, it is not sufficient to rule out a scenario with $\Delta\Neff=3$. 
For this reason we included the three scenarios in the figures. Notice that both the physical dark matter energy density and the $\sigma_8$ parameter increase as the number of DR species does, since the suppression induced in the matter power spectrum by the presence of the extra radiation species and also by their coupling to dark matter could in principle be alleviated by a larger amount of clustering dark matter as explained above.  There exists also a small degeneracy between the interaction cross section and the $\sigma_8$ parameter. This degeneracy can be easily understood in terms of Fig.~\ref{fig:matterpower} in which we show that the DM-DR interaction decreases the amplitude of dark matter fluctuations at small scales. Finally, the age of the universe in models with two or three DR species is significantly smaller than in a typical pure $\Lambda$CDM universe, since the age of the universe is inversely proportional to the amount of cold dark matter in a given cosmological scenario. 

Finally let us comment on the difference between the scenarios with constant and $\propto T^2$ DM-DR interactions. It is evident from Figs.~\ref{fig:sigma8} and~\ref{fig:sigma8b} that the isocontours very approximately transform onto each other if we associate 
\begin{equation}
15\frac{Q_0}{10^{-34}~{\rm cm}^2~{\rm MeV}^{-1}} \leftrightarrow
5\frac{Q_2}{10^{-43}~{\rm cm}^2~{\rm MeV}^{-1}} ~.
\end{equation}
This is not surprising since the DM-DR decoupling is relatively fast for both cases. 
The two cross sections are indeed similar for values of the scale factor $a_f\sim 2\times 10^4$, 
which corresponds to the epoch of matter-radiation equality. This is of course, the scale at which 
the fluctuations in the DM density can start to grow fast, unless impeded by the interactions with the DR, 
and sets the characteristic scale for the DM-DR oscillations that we can constrain.  

%%%%%%%%%%%%%%%%%%%%%%%%%%%%%%%%%%%%%%%%%%%%%%%%%%%
%%%%%%%%%%%%%%%%%%%%%%%%%%%%%%%%%%%%%%%%

%%%%%%%%%%%%%%%%%%%%%%%%%%%%%%%%%%%%%%%%

%%%%%%%%%%%%%%%%%%%%%%%%%%%%%%%%%%%%%%%%%%%%%%%%%%%

\begin{table}[!ht]
\centering
\begin{tabular}{ccc}
\hline
\hline
$\Delta \Neff $  & ($68 \%$ c.l.) & ($95 \%$ c.l.)\\
\hline
\hline
1 & $Q_0\le 5.6 $& $Q_0\le 11.8$ \\[0.1cm]
2 & $Q_0\le 2.6$& $Q_0\le 6.2$ \\[0.1cm]
3 & $Q_0\le 1.6$ & $Q_0\le 4.6$ \\[0.1cm]
\hline
\hline
\end{tabular}
\caption{Upper limits on $Q_0$ (in units of $10^{-34}$cm$^2$~MeV$^{-1}$) in different DR scenarios.}
\label{tab:Q0bounds}
\end{table}

%%%%%%%%%%%%%%%%%%%%%%%%%%%%%%%%%%%%%%%%%%%%%%%%%%%

\begin{table}[!ht]
\centering
\begin{tabular}{ccc}
\hline
\hline
$\Delta \Neff $  & ($68 \%$ c.l.) & ($95 \%$ c.l.)\\
\hline
\hline
1 & $Q_2\le 1.8 $& $Q_2\le 3.9$ \\[0.1cm]
2 & $Q_2\le 1.0$& $Q_2\le 2.7$ \\[0.1cm]
3 & $Q_2\le 0.8$ & $Q_2\le 1.9$ \\[0.1cm]
\hline
\hline
\end{tabular}
\caption{Upper limits on $Q_2$ (in units of $10^{-43}$cm$^2$~MeV$^{-1}$) in different DR scenarios.}
\label{tab:Q2bounds}
\end{table}

\section{Conclusions}
\label{sec:summary}

We have studied the phenomenological implications of models of dark matter (DM) in which the dark sector (DS) also contains additional lighter fields that interact with the DM component and constitute extra dark radiation (DR). We have been mainly inspired in models of asymmetric dark matter (ADM) that generally require complex DS extra light components into which the symmetric thermal DM abundance can efficiently annihilate. However, we tried to keep our analysis as model independent as possible and the constraints we derived can also be {applied}  to other models with DM-DR interactions regardless of the asymmetric nature of the DM component. Our description involves the splitting of the DS into light and heavy degrees of freedom ($g_\ell$ and $g_h$, respectively). While $g_\ell$ correspond to the DS degrees of freedom that will ultimately constitute the DR component, $g_h$ parametrizes the DS degrees of freedom that were relativistic at the temperature $T_d$, where the DS decoupled from the Standard Model, but that became non relativistic and heated the DR at a later time. 

The focus of our study has been put on how cosmological probes on the amount of radiation, such as measurements of the cosmic microwave background (CMB), can constrain the model. This is particularly interesting given the current preference for additional degrees of freedom in radiation displayed by the CMB, as well as the major improvement expected in this measurement with the forthcoming release of the Planck results. This provides a probe of the DS which is very complementary to more conventional searches, such as direct and indirect detection experiments, which would be challenging or absent in this type of models. We have found that the DS composition is, at present, a relatively open question with up to $\sim 20$ extra heavy degrees of freedom $g_h$ for $T_d>10$~GeV. This number is significantly suppressed once $T_d$ falls below $\sim 0.1$~GeV due to the strong reduction of SM degrees of freedom that heat the photon bath with respect to DR. Studying the impact of the forecasted Planck sensitivity we find that, if the hint for non-zero number of extra light degrees of freedom persists, the composition of the DS will be very constrained and this could provide important input for DM model building. Furthermore, if the Planck results simply put an upper limit, this limit will be strong enough to severely constrain the possible existence of light degrees of freedom in this type of models.

In addition to the effects in the very early Universe, we have also studied the possible impact of the interaction between DR and DM on structure formation. If this interaction is strong enough, the two would couple allowing the propagation of pressure waves analogous to the baryon acoustic oscillations in the baryon/photon plasma and therefore influence the galaxy power spectrum. We have studied the bounds on the interactions within the DS (in the form of DM-DR scattering) and its correlation with the number of light degrees of freedom. While the applicability of this bound is also independent of the asymmetric nature of DM, the size of the interactions required for observable effects implies too strong an annihilation of the thermal DM component to explain in this way the observed abundance. Thus, this analysis is mainly interesting for ADM scenarios in which large annihilation cross sections are desirable while the DM abundance is instead controlled by the particle-antiparticle asymmetry, as in the baryon sector. We have studied two possible forms of cross sections within the dark sector, constant and $T^2$ dependent. Using Markov Chain Monte Carlo methods, we have derived upper bounds on the strength of such interactions, which are summarized in \Tabs~\ref{tab:Q0bounds} and~\ref{tab:Q2bounds}, respectively. We have also seen that cosmological parameters, such as the age of the Universe and $\Omega_{\rm DM}h^2$, have a significant dependence on the number of interacting particles that constitute DR in these scatterings. For example, the best fit for the age of the Universe is generally lower in these scenarios as compared to a pure $\Lambda$CDM model.

\begin{acknowledgments}
We acknowledge the help of E.~Giusarma with Matlab figures and  
{partial support from the European Union FP7 ITN
INVISIBLES (Marie Curie Actions, PITN-GA-2011-289442). }
O.M. is supported by AYA2008-03531 and the Consolider Ingenio project CSD2007-00060.
\end{acknowledgments}

\providecommand{\href}[2]{#2}\begingroup\raggedright\endgroup

\end{document}